\documentclass{aa}  

\usepackage{graphicx}
\usepackage{txfonts}
\usepackage[colorlinks=true, allcolors=blue]{hyperref}
\usepackage{amsmath} 
\usepackage{CJKutf8} 

\usepackage{color}
\usepackage{placeins}

\begin{document} 

   \title{The open-source \texttt{sunbather} code: Modeling escaping planetary atmospheres and their transit spectra}

   \author{Dion Linssen \thanks{E-mail: d.c.linssen@uva.nl}
          \inst{1}
          \and
          Jim Shih (\begin{CJK*}{UTF8}{bsmi}施震\end{CJK*})\inst{1} 
          \and
          Morgan MacLeod\inst{2}
          \and
          Antonija Oklop\v{c}i\'{c}\inst{1}
          }

   \institute{Anton Pannekoek Institute for Astronomy, University of Amsterdam,
              Science Park 904, 1098 XH Amsterdam, The Netherlands
         \and
             Center for Astrophysics | Harvard \& Smithsonian, 60 Garden Street, MS-16, Cambridge, MA 02138, USA
             }

   \date{Received; accepted}

  \abstract
   {Atmospheric escape is thought to significantly influence the evolution of exoplanets, especially sub-Jupiter planets on short orbital periods. Theoretical models predict that hydrodynamic escape could erode the atmospheres of such gaseous planets, leaving only a rocky core. Deriving atmospheric mass-loss rates from observations is necessary to check these predictions. One of the ways to obtain mass-loss-rate estimates is to fit transit spectra of the 10830~Å helium or UV metal lines with Parker wind models. We aim to provide the community with a tool that enables this type of analysis, and present \texttt{sunbather}, an open-source Python code that can be used to model escaping exoplanet atmospheres and their transit spectra. \texttt{sunbather} incorporates the Parker wind code \texttt{p-winds} and the photoionization code \texttt{Cloudy}, with the ability to calculate any currently known spectral tracer, using an arbitrary atmospheric composition. With \texttt{sunbather}, we investigate how the atmospheric structure of a generic hot-Neptune planet depends on metallicity. We find that the mass-loss rate drops by roughly one order of magnitude as we increase the metallicity from solar to 50 times solar. Line cooling by metal species is already important for a solar composition, and is even more so at higher metallicity. We then demonstrate how \texttt{sunbather} can be used to interpret observations of spectral lines that form in the upper atmosphere. We fit the observed helium spectrum of the mini-Neptune TOI-2134~b and show how, even for helium data, the inferred mass-loss rate can change by a factor of up to three, depending on the assumed metallicity.}


   \maketitle

\section{Introduction} \label{sec:introduction}
Recent advances in the field of exoplanet research have identified a lack of sub-Jovian planets, as well as a bimodal distribution of the radii of sub-Neptune-sized planets, both at short orbital periods \citep{szabo_short-period_2011, beauge_emerging_2013, owen_kepler_2013, lundkvist_hot_2016, fulton_california-_2017}. Atmospheric escape is thought to play a major role in sculpting these features. Theoretical works aiming to explain the observed demographic characteristics in this way typically prescribe analytical mass-loss-rate estimates based on the bulk parameters of planets \citep[e.g.,][]{owen_evaporation_2017, ginzburg_core-powered_2018}. However, the relative importance of atmospheric escape compared to other proposed mechanisms such as formation and migration is still a matter of debate \citep[e.g.,][]{owen_photoevaporation_2018, lee_creating_2022, burn_radius_2024}. The same is true for the physical details of the atmospheric escape process, which can be due to photoevaporation or core-powered mass loss, or indeed both \citep[e.g.,][]{owen_mapping_2024}, with possible contributions from giant impacts \citep[e.g.,][]{chance_signatures_2022, zhong_impact-driven_2024}. Therefore, obtaining mass-loss rates that are based on observational measurements for a large sample of individual planets would allow us to benchmark the theoretical predictions and test the atmospheric escape hypothesis \citep{vissapragada_upper_2022}. 

Observational evidence of atmospheric escape for individual planets is gathered through transit spectroscopy of specific spectral lines. In the early days of atmospheric characterization, this consisted of UV observations of the hydrogen Lyman-$\alpha$ line and lines from metals such as oxygen, carbon, and magnesium \citep[e.g.,][]{vidal-madjar_extended_2003, vidal-madjar_detection_2004, fossati_metals_2010}. In more recent years, the 10,830~Å line of metastable helium has additionally been identified as a tracer of atmospheric escape \citep{spake_helium_2018, oklopcic_new_2018, nortmann_ground-based_2018, allart_spectrally_2018}. Consequently, the number of atmospheric escape detections has rapidly grown to a few dozen \citep{dos_santos_observations_2023}, enabling the search for trends at the population level \citep[e.g.,][]{allart_homogeneous_2023, bennett_nondetection_2023, krishnamurthy_helium_2024}. 

Based on the specific spectral line considered, as well as the quality and complexity of the data, different types of models may be needed for their interpretation. Models ranging in complexity from a relatively simple 1D isothermal Parker wind \citep[e.g.,][]{mansfield_detection_2018} to expensive 3D coupled hydrodynamics-thermochemistry simulations \citep[e.g.,][]{wang_metastable_2021} have been applied in the literature. With the goal being to infer a mass-loss rate from a spectral line measurement, the isothermal Parker wind prescription has been widely used because it treats the mass-loss rate as a free parameter \citep[e.g.,][]{mansfield_detection_2018, lampon_modelling_2020, lampon_evidence_2021, lampon_characterisation_2023, kasper_nondetection_2020, palle_he_2020,  vissapragada_constraints_2020, vissapragada_search_2021, vissapragada_upper_2022, zhang_no_2021, zhang_outflowing_2023, zhang_detection_2023, krishnamurthy_nondetection_2021, krishnamurthy_absence_2023, paragas_metastable_2021, czesla_h_2022, dos_santos_p-winds_2022, kirk_kecknirspec_2022, spake_non-detection_2022, allart_homogeneous_2023, bennett_nondetection_2023, gaidos_planetesimals_2023, orell-miquel_confirmation_2023, gully-santiago_large_2024, perez-gonzalez_detection_2024}. This prompted \citet{dos_santos_p-winds_2022} to release \texttt{p-winds}, an open-source implementation of the isothermal Parker wind model that includes a module to produce metastable helium transit spectra. Although the isothermal Parker wind model has thus far been mainly used for interpreting helium observations, it need not be limited to this spectral line, and the newest version of \texttt{p-winds} includes the ability to make carbon and oxygen spectra \citep{dos_santos_hydrodynamic_2023}. Indeed, in addition to the upsurge in the number of helium studies in this regard, there have been several recent detections of escaping metals in the UV \citep[e.g.,][]{sing_hubble_2019, cubillos_near-ultraviolet_2020, cubillos_hubble_2023, garcia_munoz_heavy_2021, ben-jaffel_signatures_2022, sreejith_cute_2023}, which provide highly complementary data that would benefit from similar modeling strategies.

Here, we aim to address this need and present \texttt{sunbather}\footnote{\label{fn:github}\url{www.github.com/dlinssen/sunbather}}, an open-source code to model the structure and transit spectra of escaping exoplanet atmospheres. The code couples \texttt{p-winds} to the detailed nonlocal thermodynamic equilibrium (NLTE) photoionization code \texttt{Cloudy} \citep{ferland_cloudy_1998, ferland_2017_2017, chatzikos_2023_2023}. This enables the calculation of atmospheres of  arbitrary composition and of any currently known spectral line tracing escape \citep{linssen_expanding_2023}. Starting from the isothermal Parker wind as produced by \texttt{p-winds}, \texttt{sunbather} calculates a more realistic nonisothermal temperature profile, which can have important consequences for the predicted transit spectrum. As shown in \citet{linssen_constraining_2022}, the nonisothermal profile can additionally be used to restrain the model parameter space and yield more stringent mass-loss-rate constraints.

This paper is organized as follows. In Sect. \ref{sec:code}, we describe the theory behind the model, and explain how it is implemented in three main Python modules. In Sect. \ref{sec:metallicity}, we use \texttt{sunbather} for a theoretical study of the effects of high metallicity on the upper atmospheric structure and transit signatures of a sub-Neptune planet. In Sect. \ref{sec:constraining}, we demonstrate how \texttt{sunbather} can be used to constrain mass-loss rates from spectral-line observations. In Sect. \ref{sec:summary}, we summarize our findings. A reproduction package for all of the results of this manuscript is available online\footnote{\label{fn:zenodo}\url{https://doi.org/10.5281/zenodo.11202780}}.

\section{The \texttt{sunbather} code} \label{sec:code}
We model an exoplanet atmosphere that is escaping hydrodynamically. In this escape regime, gas particles obtain an average internal energy comparable to that needed to escape from the planet \citep{gronoff_atmospheric_2020}. Hence, thermal expansion of the atmosphere drives a bulk outflow, requiring that the gas is collisional up until the sonic point \citep{owen_atmospheric_2019}. The thermal energy required can either be supplied by stellar high-energy radiation (called photoevaporation) or the planet's interior luminosity (called core-powered mass-loss). Hydrodynamic escape is expected to be the dominant mechanism for highly irradiated planets.

Following \citet{murray-clay_atmospheric_2009}, the Euler equations describing a steady-state hydrodynamic outflow in the planetary frame in 1D are as follows. Mass continuity can be written as
\begin{equation} \label{eq:mass_cons}
    \frac{\partial}{\partial r}(r^2 \rho v) = 0,
\end{equation}
where $r$ is the distance to the planet center, $\rho$ the density and $v$ the velocity. Momentum conservation implies
\begin{equation} \label{eq:momentum_cons}
    \rho v \frac{\partial v}{\partial r} + \frac{\partial P}{\partial r} +\frac{G M_p \rho}{r^2} + \frac{3 G M_* \rho r}{a^3}= 0,
\end{equation}
where $P$ is pressure, $G$ the gravitational constant, $M_p$ the planet mass, $M_*$ the stellar mass, and $a$ the semi-major axis. The last term on the left-hand side is the stellar tidal gravity term. It can be optionally excluded in \texttt{sunbather}. Energy balance reads
\begin{equation} \label{eq:energy_cons}
    - \rho v \frac{\partial}{\partial r}\Bigg[ \frac{k T}{(\gamma -1) \mu}\Bigg] + \frac{k T v}{\mu} \frac{\partial \rho}{\partial r} + \Gamma + \Lambda = 0,
\end{equation}
where $k$ is the Boltzmann constant, $T$ the temperature, $\gamma=5/3$ the adiabatic index for a perfect gas, $\mu$ the mean particle mass, $\Gamma$ the radiative heating rate, and $\Lambda$ the radiative cooling rate. Solving this system of equations results in a ``self-consistent'' solution to the problem. One obtains the radial density, velocity, and temperature profiles, as well as the corresponding mass-loss rate of the planet. This exercise is performed by various open-source codes, such as \texttt{TPCI} \citep{salz_TPCI_2015}, \texttt{ATES} \citep{caldiroli_irradiation-driven_2021}, and \texttt{AIOLOS} \citep{schulik_span_2023}. A useful asset of these self-consistent codes is that they have predictive power of the atmospheric mass-loss rate. However, the associated predicted transit spectrum may be inconsistent with observational data.

The isothermal Parker wind \citep{parker_dynamics_1958, lamers_introduction_1999} is an alternative type of model that parametrizes the mass-loss rate and temperature of the atmosphere. The model thus does not aim to include all physics and solve the atmospheric structure from first principle. Rather, it assumes ignorance of some parameters and aims to infer those from data. Mathematically, the isothermal Parker wind description comes down to ignoring Eq. \ref{eq:energy_cons}. Instead, the mass and momentum equations (Eqs. \ref{eq:mass_cons} and \ref{eq:momentum_cons}) are solved while assuming a constant sound speed
\begin{equation} \label{eq:sound_speed}
v_s=\sqrt{\frac{kT}{\mu}}.
\end{equation}
One then obtains the radial density and velocity profiles \citep[see e.g.,][]{lamers_introduction_1999}, and a transit spectrum can be calculated. Using this model, an observed transit spectrum can virtually always be matched with some values of the mass-loss rate and temperature, allowing these parameters to be retrieved from observations. This is especially useful for the mass-loss rate, which is the quantity we are often most interested in.

A common practice in the literature is to fit isothermal Parker wind models with different mass-loss rates and temperatures to an observed transit spectrum \citep[e.g.,][]{lampon_modelling_2020, vissapragada_constraints_2020, allart_homogeneous_2023}. However, the best-fit models often show a degeneracy between the temperature and mass-loss rate, making it difficult to constrain them. Especially retrievals on data with low spectral resolution suffer from this degeneracy. It is less of a problem in higher resolution spectra, since the temperature can be constrained from the spectral line width. 

In low-resolution spectra, an external constraint on the temperature of the atmosphere can help to break the degeneracy and hence constrain the mass-loss rate \citep{vissapragada_maximum_2022, linssen_constraining_2022}. In \texttt{sunbather}, we provide a way to find a sensible temperature of the Parker wind and hence allow constraining the mass-loss rate. After having calculated the density and velocity profiles of the isothermal Parker wind, we plug them into Eq. \ref{eq:energy_cons} and solve for a new temperature profile that is nonisothermal. The density, velocity, and temperature profiles are thus not entirely self-consistent, given that the former two are calculated assuming an isothermal profile. That being said, exactly this fact can be leveraged to estimate a sensible Parker wind temperature. If the isothermal and nonisothermal temperature profiles are similar, the atmospheric structure and hence temperature is relatively self-consistent. Limiting the model parameter space to only those Parker wind models (for example by using a prior) means that effectively the mass-loss rate is the only free parameter left to fit to the data. We first demonstrated this method in \citet{linssen_constraining_2022}. A second reason to calculate a nonisothermal temperature profile is that it should simply be more realistic than an assumed isothermal profile. Apart from affecting the atmospheric structure and line widths, the temperature also affects the chemical state of gas (i.e.,the ionization state and atomic level populations) and hence the transit spectrum. 

\texttt{sunbather} includes the 30 lightest elements (up until zinc) and allows for an arbitrary composition. The inclusion of metal species compared to a pure hydrogen/helium composition manifests itself in the outflow properties and transit spectrum in various ways, which we explore in Sect. \ref{sec:metallicity}. The composition of the atmosphere is kept constant with radius in the atmosphere. Heavy metal particles are assumed to be dragged along efficiently in the outflow so that there is no mass fractionation. This assumption breaks down when the mass-loss rate is low and only the lighter elements are allowed to escape \citep{hunten_mass_1987}. It is up to the user of the code to consider whether each species is expected to escape and if the chosen composition is sensible. This can for example be estimated using the crossover mass \citep{hunten_mass_1987}. The cases in which heavier particles either escape efficiently or do not escape at all can be modeled well by choosing the appropriate (constant) composition. Only the intermediate case in which heavy particles do escape, but at a lower rate than the lighter elements, cannot be appropriately modeled as it requires abundance patterns that change with radius in the atmosphere.

The transit spectrum can be calculated over any wavelength range between 1~Å and 10~$\mathrm{\mu}$m. It aims to include all atomic and ionic spectral lines present in the NIST Atomic Spectra Database\footnote{\url{https://www.nist.gov/pml/atomic-spectra-database}}. Not all of these lines can be calculated (see Sect. \ref{sec:RT} for more details), but the strongest lines are included. To quantify, 29,530 spectral lines are included, equal to 40\% of the lines in the NIST database. Sources of continuum opacity, as well as molecular absorption lines cannot be calculated in the current version of \texttt{sunbather}, but molecules are also expected to be dissociated in the outflow. Our model is 1D and we assume spherical symmetry, and so we cannot produce asymmetric spectral line shapes or light curves. Spectral lines will always be symmetric around their rest-frame wavelength.

\begin{figure*}
   \centering
   \includegraphics[width=\hsize]{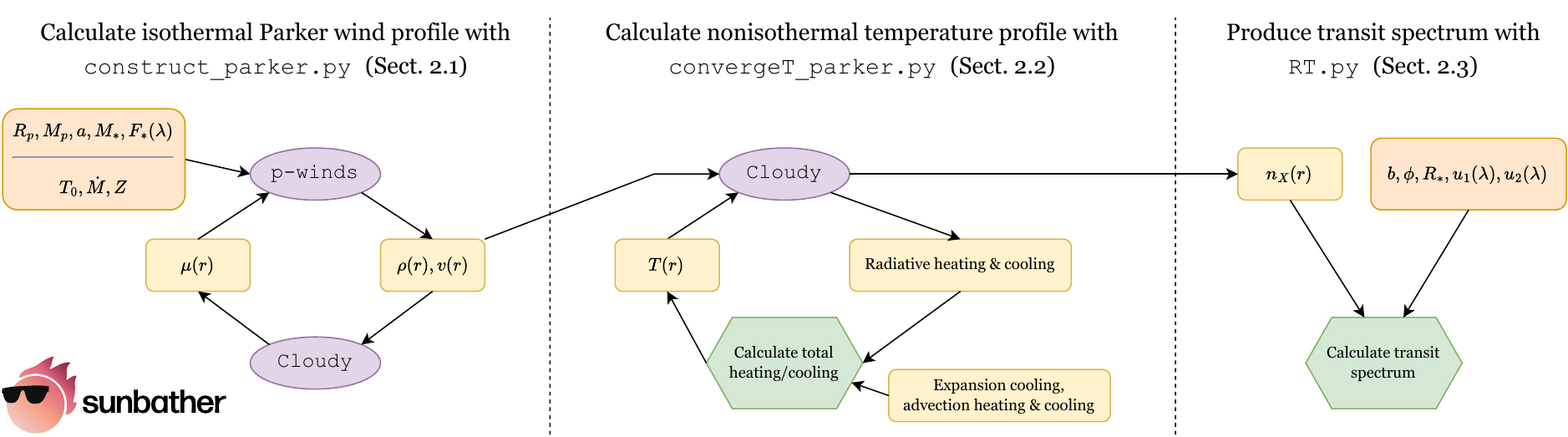}
      \caption{Simplified visualization of the modeling setup of \texttt{sunbather}. The variables are as follows: $R_p$ and $M_p$ are the planet radius and mass, $a$ is the orbital distance (only circular orbits are possible), $M_*$ and $F_*(\lambda)$ are the stellar mass and spectral energy distribution, $T_0$ and $\dot{M}$ are the (constant) temperature and mass-loss rate of the Parker wind, $Z$ denotes the atmospheric bulk composition, $\mu(r)$, $\rho(r)$, $v(r),$ and $T(r)$ are the atmospheric mean particle mass, density, velocity, and (nonconstant) temperature profiles, $n_X(r)$ represents the number density of any ``chemical species'' (i.e., any energy level of an atom or ion), $b$ and $\phi$ are the planet transit impact parameter and orbital phase, and $R_*$, $u_1(\lambda)$ and $u_2(\lambda)$ are the stellar radius and quadratic limb-darkening parameters.}
         \label{fig:sunbather_chart}
   \end{figure*}

We now describe the setup of \texttt{sunbather} in more detail. This is divided into three sections that correspond to the modules of \texttt{sunbather} in which they are implemented. Fig. \ref{fig:sunbather_chart} presents a simplified visualization of the computational scheme.

\subsection{Calculating the density and velocity structure: The \texttt{construct\_parker} module} \label{sec:construct}
In this module, we calculate isothermal Parker wind density and velocity profiles using a combination of \texttt{p-winds} \citep{dos_santos_p-winds_2022} and \texttt{Cloudy} \citep{ferland_cloudy_1998, ferland_2017_2017, chatzikos_2023_2023}. Despite what the name suggests, the wind is not strictly isothermal in this formalism. Instead, the sound speed and thus the ratio of $T(r)/\mu(r)$ is constant. $\mu(r)$ depends on the composition of the atmosphere (which we keep constant with radius), but also the ionization fractions, which do change with radius. $\mu$ therefore typically varies by a factor $\lesssim2$ throughout the atmosphere. If one calculates an appropriately averaged value $\bar{\mu}$, a corresponding averaged isothermal value $T_0$ can also be assigned \citep[Appendix A of][]{lampon_modelling_2020}.

We use \texttt{p-winds} to calculate the density and velocity profiles of the atmosphere. This means it solves Eqs. \ref{eq:mass_cons} and \ref{eq:momentum_cons}; we refer the reader to \citet{dos_santos_p-winds_2022} for the details of this calculation. To run, \texttt{p-winds} needs the planet mass $M_p$, the planet radius $R_p$, and the stellar spectrum $F_*(\lambda)$ at the planet orbital distance $a$ as input. Additionally, the three free parameters of the model are the isothermal temperature $T_0$, the bulk mass-loss rate $\dot{M}$, and the composition of the atmosphere. These parameters cannot be predicted by the model - they have to be assumed. In \texttt{p-winds}, the atmosphere is assumed to be hydrogen/helium only, meaning that choosing the composition is equivalent to choosing a hydrogen abundance fraction $f_H$. The code calculates the ionization structures of hydrogen and helium, resulting in a $\mu(r)$ profile and associated $\bar{\mu}$. Together with $T_0$, $\bar{\mu}$ determines the speed of sound, which is needed for the calculation of the density and velocity profiles. If the user of \texttt{sunbather} chooses an atmospheric composition that is hydrogen/helium only, \texttt{p-winds} can be used standalone to obtain the density and velocity profiles. However, when the chosen atmospheric composition includes metals, the $\mu(r)$-structure as calculated by \texttt{p-winds} will not be correct. We therefore use \texttt{Cloudy} to calculate $\mu(r)$ when metals are included. 

\texttt{Cloudy} is a 1D NLTE photoionization code applicable to many different astrophysical problems. Currently, versions \texttt{v17.02} and \texttt{v23.01} are supported by \texttt{sunbather}, and the user may decide which version to install (no significant differences between the two versions were found). \texttt{Cloudy} calculates microphysical processes inside a gas cloud irradiated by a light source, predicting the ionization and energy level populations, as well as the emerging spectrum. It includes extensive databases of energy levels, transition probabilities, and collision rates for different molecular, atomic, and ionic species. This makes it very suitable for our purpose of predicting the physical state of upper exoplanet atmospheres, which due to their low densities are expected to show a complex chemical state governed by NLTE effects. In recent years, \texttt{Cloudy} has been increasingly applied to the study of exoplanet atmospheres \citep[e.g.,][]{salz_simulating_2016, fossati_data-driven_2020, fossati_non-local_2021, fossati_gaps_2023, turner_detection_2020, young_non-local_2020, young_searching_2024, zhang_no_2021, zhang_escaping_2022, linssen_constraining_2022, linssen_expanding_2023, kubyshkina_precise_2024}. 

The isothermal Parker wind density and velocity profiles are solved in an iterative manner when metals are included (visualized in the left part of Fig. \ref{fig:sunbather_chart}). We start by running \texttt{p-winds} to calculate the density and velocity profile with hydrogen and helium only. We then simulate this density profile with \texttt{Cloudy}, while fixing an isothermal temperature profile $T(r)=T_0$, and using the full composition that includes metals. \texttt{Cloudy} calculates - amongst other things - the ionization profiles of all elements and the corresponding free electron number density profile. If we ignore molecules, we only need the electron density to calculate the mean particle mass as
\begin{equation} \label{eq:mu}
    \mu(r) = \frac{\sum_s X_s m_s}{1 + \frac{n_e(r)}{\rho(r)} \sum_s X_s m_s},
\end{equation}
where $X_s$ and $m_s$ are the abundance and mass of species $s$, respectively, and $n_e$ is the free electron number density. The abundances are expressed as a dimensionless fraction and obey $\sum_s X_s = 1$. We use Eq. A.3 from \citet{lampon_modelling_2020} to find $\bar{\mu}$ from $\mu(r)$. This completes the first iteration and we start the second. We run \texttt{p-winds} again and use the new $\bar{\mu}$ value, while bypassing its own calculation of $\bar{\mu}$. This results in a new density and velocity profile. We simulate the density profile with \texttt{Cloudy} to obtain a new $\bar{\mu}$, completing the second iteration. We repeat these steps until $\bar{\mu}$ is converged to $(|\bar{\mu}_{i}-\bar{\mu}_{i-1}|)/\bar{\mu}_{i-1}<0.01$, where $i$ is the iteration number. \texttt{sunbather} also stops if convergence is not reached after seven iterations (the user can change the convergence threshold and maximum iteration number).

The runtime of this module depends on the composition. For a pure hydrogen/helium composition, \texttt{p-winds} can calculate the density and velocity profiles in a second. When the \texttt{p-winds/Cloudy} iterative scheme is applied, the runtime is of order ten minutes. One could also still invoke the iterative scheme when calculating a pure hydrogen/helium composition, as \texttt{Cloudy} may report a different $\mu(r)$-structure from \texttt{p-winds}.

\subsection{Calculating the temperature structure: The \texttt{convergeT\_parker} module} \label{sec:convergeT}
In this module, we calculate a nonisothermal temperature structure using \texttt{Cloudy}. This means we now introduce Eq. \ref{eq:energy_cons} and solve it for $T(r)$ while using $\rho(r)$ and $v(r)$ that were calculated before under the isothermal assumption. The four terms of Eq. \ref{eq:energy_cons} represent advection, expansion, radiative heating, and radiative cooling, respectively. The advection and expansion terms depend explicitly on the temperature profile. The radiative heating and cooling terms take a few minutes to calculate using \texttt{Cloudy} and implicitly depend on the temperature in a nontrivial manner. Therefore, we cannot instantly solve Eq. \ref{eq:energy_cons} for the temperature profile, but use an iterative algorithm instead. The middle part of Fig. \ref{fig:sunbather_chart} shows a simplified schematic of the iterative algorithm, and Fig. \ref{fig:convergence} shows an example case.

We start by running the density profile through \texttt{Cloudy} with an initial temperature profile. The user of \texttt{sunbather} can choose whether this initial temperature profile is
\begin{enumerate}[i)]
  \item the isothermal profile assumed when calculating the density and velocity profiles,
  \item the temperature profile of a previously calculated model with a similar $T_0$ and $\dot{M}$,
  \item unspecified, in which case \texttt{Cloudy} will calculate it assuming radiative equilibrium.
\end{enumerate}
The chosen initial temperature profile does not affect the final nonisothermal profile, but it does affect the number of iterations needed to find the solution. The output of this first \texttt{Cloudy} simulation is $\Gamma(r)$, $\Lambda(r)$, $\mu(r)$ (after applying Eq. \ref{eq:mu}), as well as $T(r)$ if option iii) was chosen. For this initial temperature profile, Eq. \ref{eq:energy_cons} will not yet be satisfied. Therefore, we guess a new temperature structure and run \texttt{Cloudy} again to obtain new $\Gamma(r)$, $\Lambda(r)$, and $\mu(r)$ profiles. This is repeated until Eq. \ref{eq:energy_cons} is respected. We use two complementary algorithms to guess the temperature structure of the next iteration, which we  describe in the following two subsections. 

\subsubsection{Algorithm 1: ``Relaxing'' the temperature profile} \label{sec:algorithm1}
\begin{figure}
   \centering
   \includegraphics[width=\linewidth]{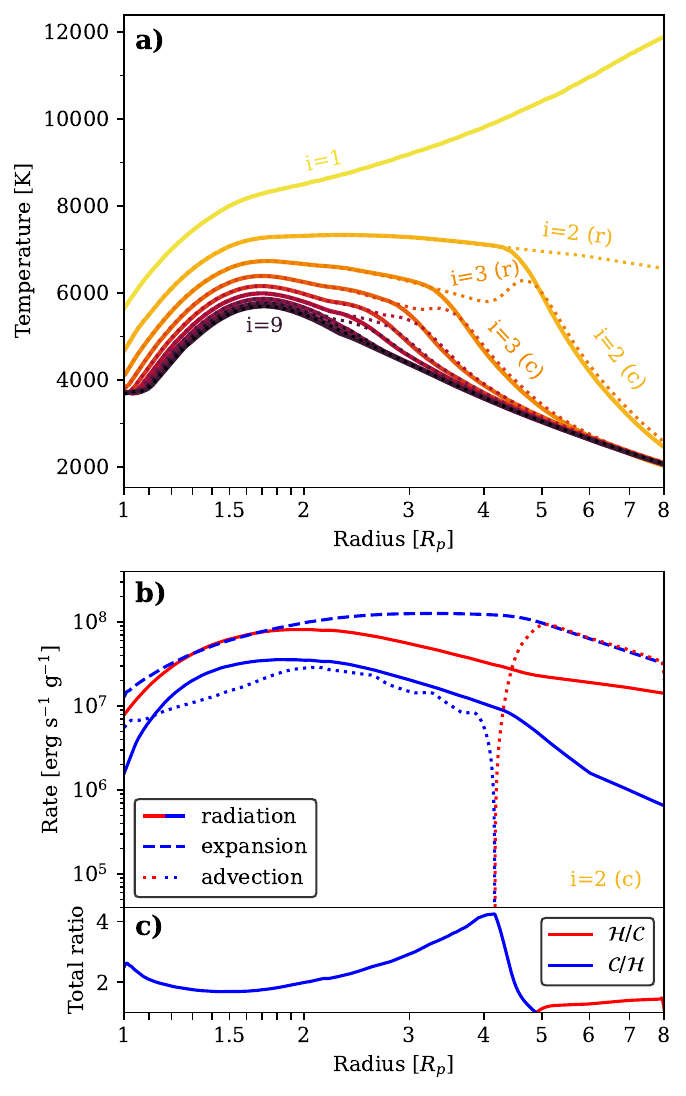}
      \caption{Iterative method to solve for the nonisothermal temperature profile. {\bf a)} Temperature profiles of the different iterations $i$. Convergence is reached after nine iterations. Upon each iteration, two algorithms are applied to guess the temperature profile. Dotted lines show the profile based on the relax algorithm of Sect. \ref{sec:algorithm1} and are labeled (r). Solid lines then show the profile after the construct algorithm of Sect. \ref{sec:algorithm2} has also been applied and are labeled (c). {\bf b)} The different heating and cooling rates corresponding to the temperature profile of the second iteration. {\bf c)} Ratio of the total heating rate to total cooling rate, for the rates shown in panel (b). If the ratio is $<f_c=1.1$ everywhere, the temperature profile is considered converged. Based on the $\mathcal{H}/\mathcal{C}$ and $\mathcal{C}/\mathcal{H}$ ratios shown here, the guessed $i$=3 (r) profile will be cooler at $r<5R_p$ and slightly hotter at $r>5R_p$, as confirmed in panel (a).}
         \label{fig:convergence}
   \end{figure}

In this algorithm, we guess a new temperature profile based on the heating/cooling inequality of the last iteration. We aim to find the temperature profile for which the total heating and cooling rate are equal and hence their ratio equal to 1, as that implies we have solved Eq. \ref{eq:energy_cons}. After running \texttt{Cloudy} with the temperature profile of the last iteration, we have all the quantities needed to calculate the four different terms of Eq. \ref{eq:energy_cons}. The advection term can be positive (heating) or negative (cooling), depending on the slope of $T(r)/\mu(r)$, while expansion and $\Lambda(r)$ are always negative and $\Gamma(r)$ is always positive. Fig. \ref{fig:convergence}b shows an example of these rates for the temperature profile of the second iteration. We sum the different heating and cooling rates into a total heating rate $\mathcal{H}(r)$ and total cooling rate $\mathcal{C}(r)$ (which we define as positive here) as a function of radius. Fig. \ref{fig:convergence}c shows the ratio $\mathcal{H}(r)/\mathcal{C}(r)$ as a red line if heating is stronger or $\mathcal{C}(r)/\mathcal{H}(r)$ as a blue line if cooling is stronger. Based on this ratio, we guess a new temperature structure, assigning a higher temperature to radii where $\mathcal{H}>\mathcal{C}$, and a lower temperature to radii where $\mathcal{C}>\mathcal{H}$. Specifically,
\begin{equation}
T_{i+1}(r) =
    \begin{cases}
        T_{i}(r) \dfrac{1}{1-\delta_i(r) \cdot \mathrm{log}_{10}\left(\frac{\mathcal{H}(r)}{\mathcal{C}(r)}\right)} & \text{if } \mathcal{H}(r) > \mathcal{C}(r)\\ \\
        T_{i}(r) \left[1-\delta_i(r) \cdot \mathrm{log}_{10}\left(\frac{\mathcal{C}(r)}{\mathcal{H}(r)}\right)\right] & \text{if } \mathcal{H}(r) < \mathcal{C}(r),
    \end{cases}
\end{equation}
where $i$ is the iteration number and $\delta(r)$ is a scaling factor that sets how large the temperature changes are between iterations. In the first iteration, $\delta_1(r)=0.3$ at all radii. In some cases, the changes to the temperature structure are too aggressive and $T(r)$ will fluctuate up and down between iterations. We check for radial bins in which such temperature fluctuations happen, and we lower the scaling factor there as $\delta_{i+1}(r)=2/3 \cdot \delta_i(r)$. This ensures smaller changes in temperature between iterations, preventing fluctuations and allowing the temperature structure to converge. $T_{i+1}/T_i$ is furthermore bounded by 0.5 and 2 to prevent extremely large jumps in temperature. In Fig. \ref{fig:convergence}a, the dotted line labeled ``iteration 3'' shows the newly guessed temperature profile after applying the algorithm described in this section. We do not immediately start the third iteration with this profile, however. First, we apply a second, complementary algorithm described in the next subsection to refine the guessed temperature profile of iteration 3 further.

\subsubsection{Algorithm 2: ``Constructing'' the temperature profile} \label{sec:algorithm2}
In this algorithm, we further modify the guessed temperature from the relaxation algorithm described in the previous section, before starting the next iteration. Contrary to the relaxation algorithm, we here do not use the temperature profile of the last iteration. Instead, we use only $\mu(r)$, $\Gamma(r)$ and $\Lambda(r)$ of the last iteration as given by \texttt{Cloudy}. Together with $\rho(r)$ and $v(r)$, we can solve Eq. \ref{eq:energy_cons} for $T(r)$, which is the only remaining variable. The reason that this does not immediately yield the correct temperature structure, is that $\Gamma$ and $\Lambda$ implicitly depend on the temperature, meaning that the $\Gamma(r,T)$ and $\Lambda(r,T)$ values from the last iteration are not consistent anymore with the newly constructed $T_{i+1}(r)$. Iterations are therefore still required to find the converged temperature structure. 

This algorithm will not work well in parts of the atmosphere where the dominant heating and cooling terms are the radiation rates. This is because we solve Eq. \ref{eq:energy_cons} while keeping the radiation rates fixed at those from the last iteration, such that only the expansion and advection terms can change. If the latter are not the dominant terms, the algorithm will make huge adjustments to the temperature structure in order to change the expansion and advection terms enough to achieve heating/cooling balance. However, smaller changes to the temperature profile would actually already lead to energy balance, since $\Gamma(r,T)$ and $\Lambda(r,T)$ would change accordingly as well. Oppositely, this algorithm works well and achieves convergence quickly when radiation is relatively unimportant. In this case, the radiation rates are still fixed at those from the last iteration and will again not be fully consistent with the newly constructed temperature structure. However, this is not important for the total heating/cooling balance, which is dominated by advection and/or expansion. For these reasons, we always employ the relaxation algorithm of Sect. \ref{sec:algorithm1} first, and afterwards only employ the construction algorithm of this section over the region of the atmosphere where radiation is relatively unimportant (provided such a region exists). 

We construct the new temperature profile as follows. Given a starting value $T(r_{start})$ specified at one particular radial bin $r_{start}$, we calculate $T$ in the next radial bin by minimizing the left-hand side of Eq. \ref{eq:energy_cons} with respect to $T$ (using the \texttt{minimize\_scalar} function from the \texttt{scipy} library). We can then calculate $T$ in the following radial bin and proceed in this way until the full profile is found. The temperature profile constructed with this method is very sensitive to $T(r_{start})$, and therefore we only apply the algorithm if there is a starting point where the temperature profile is already almost converged, through the use of the relaxation algorithm described in Sect. \ref{sec:algorithm1}. Additionally, we only apply the algorithm if the whole region of the atmosphere that we construct (i.e., $r>r_{start}$) is dominated by expansion or advection, for the reasons discussed above. The exact criterion to find a suitable $r_{start}$ is somewhat complex and we refer the interested reader to the \texttt{sunbather} source code. Roughly speaking, we require $\mathcal{H}(r_{start})/\mathcal{C}(r_{start})$ or $\mathcal{C}(r_{start})/\mathcal{H}(r_{start})$ (whichever is higher) to be smaller than 1.3, but this value is allowed to be higher if the contributions from radiative heating and cooling are sufficiently low.

\subsubsection{Reaching convergence}
Our complete method to calculate the nonisothermal temperature profile consists of iteratively running the temperature profile through \texttt{Cloudy}, updating the temperature profile using the described algorithms, and radially smoothing the temperature profile over small scales. We smooth the temperature profile mainly because the advection rate shows small discrete jumps that propagate to jumps in the newly guessed temperature profile. These jumps arise from the dependence of the advection rate on the temperature gradient and \texttt{Cloudy}'s limited precision in reporting the temperature. Without smoothing, small discrete jumps in the temperature profile would lead to much larger changes in the advection rate in the next iteration, quickly destabilizing the algorithm. We consider the temperature profile converged when $1/f_c < \mathcal{H}(r)/\mathcal{C}(r)<f_c$ at all radii. The default value of the convergence threshold $f_c$ is 1.1 and can be changed by the user of the code. In some cases, the smoothing of the temperature profile prevents it from reaching a convergence factor of $f_c$, even though the profile does not change substantially anymore between iterations. These solutions are thus correct up to the level of the smoothing (typically $\lesssim 50$~K), and we consider them valid as well. The runtime of this module is on the order of ten minutes.

We finally clarify that the advection term of Eq. \ref{eq:energy_cons} only accounts for the advective heat flow. In principle, the advective particle flow should also lead to the transport of atoms from the deeper layers of the atmosphere to the ionic upper layers of the atmosphere, thereby influencing the degree of ionization. This would in turn affect the mean particle mass that we calculate in Eq. \ref{eq:mu}, as well as the radiative heating and cooling rates through different ionization and recombination rates. Properly including the advective velocity component in \texttt{Cloudy} would require editing the source code \citep{salz_TPCI_2015} and drastically increase its computation time, and so we ignore these second-order effects on $\mu$, $\Gamma$, and $\Lambda$. 

\subsection{Calculating the transmission spectrum: The \texttt{RT} module} \label{sec:RT}
In this module, we use the \texttt{Cloudy} output to calculate the transmission spectrum. We do not use the spectrum predicted by \texttt{Cloudy}. \texttt{Cloudy} is a hydrostatic code and thus does not include spectral line broadening due to the outflow velocity. Furthermore, instead of one simulation along the substellar direction, multiple simulations at different impact parameters from the planet would be needed to produce the correct transit geometry \citep[see e.g.,][]{young_non-local_2020}. Therefore, we instead postprocess the chemical state of the atmosphere that \texttt{Cloudy} calculates with a ray-tracing calculation as follows.

The user provides a wavelength region and a set of atomic and ionic species for which to calculate the transit spectrum. For each species, \texttt{sunbather} includes a downloaded table of all spectral line coefficients of the NIST database. This table is filtered to only those lines that fall within the requested wavelength range. Each absorption line corresponds to a transition from a lower energy level of an atom/ion to a higher energy level. To calculate a spectral line, the number density of the lower energy level as a function of radius in the atmosphere is needed. For a given line in the NIST database, \texttt{sunbather} therefore first tries to identify if the lower energy level can be matched to any of the energy levels used by \texttt{Cloudy}. This is not always possible, because we use the default \texttt{Cloudy} energy level selection, which for most species includes the 15 lowest levels. Additionally, \texttt{Cloudy} will sometimes ``fold'' multiple energy levels together into one, or include energy levels that simply cannot be matched to the levels reported by NIST. In those cases, the spectral line cannot be calculated. In practice, however, the strongest lines in a transit spectrum originate from the lowest few energy levels of an atom or ion, which are typically included in the default \texttt{Cloudy} selection. Therefore, we choose to not increase the number of energy levels included in \texttt{Cloudy} beyond the default selection, as this would increase the computation time. For the energy levels that can be matched to NIST, \texttt{Cloudy} provides the number density of that energy level as a function of radius in the atmosphere.

We then project the 1D number density of the energy level $n_X(r)$, temperature $T(r),$ and velocity $v(r)$ profile onto a 2D semi-circle with the $x$-coordinate along the direction of the stellar light rays and the $y$-coordinate indicating the impact parameter of the light rays with respect to the planet surface. The velocity is directed radially outward, but only the $v_x$-component affects the Doppler shift of the observed line. By default, \texttt{sunbather} uses 100 rays, but this number can be changed if higher or lower precision is needed. The code then loops over each individual spectral line and calculates the cumulative optical depth of each of the 100 rays as a function of frequency $\nu$ as
\begin{equation}
    \tau_y(\nu) = \int_x n_X(x,y) \sigma_0 \Phi(\Delta \nu) \mathrm{d}x,
\end{equation}
where $\sigma_0=2.654 \times 10^{-2} \times f_{12}$ is the absorption cross-section of the given spectral line. $f_{12}$ is the oscillator strength of the line given by the NIST database. $\Phi$ is the Voigt line profile, which is a convolution of a Lorentzian line profile due to natural broadening and a Gaussian line profile due to thermal broadening. The Lorentzian profile has a (constant) half-width at half-maximum of $\gamma=A_{21}/4\pi$, where $A_{21}$ is the Einstein coefficient of the line given by the NIST database. The Gaussian profile has a (temperature-dependent) half-width at half-maximum of 
\begin{equation}
    \alpha = \sqrt{\frac{2 \mathrm{ln}2 \; k T(x,y)}{m_X}} \frac{\nu_0}{c},
\end{equation}
where $m_X$ is the mass of the absorbing species and $v_0$ is the rest-frame frequency of the line (i.e., the line center). The line profile is dominated by thermal broadening around the line center, but natural broadening can dominate the absorption in the far wings of the line. If the Voigt profile itself is centered around zero, it is evaluated at 
\begin{equation}
\Delta \nu = (\nu - \nu_0) - \frac{v_x(x,y)}{c}\nu_0
\end{equation}
in order to take into account the deviation from the line center at the current frequency, as well as the Doppler shift due to the outflow velocity. We then sum the $\tau_y(\nu)$ contributions of all spectral lines together. 

The last step is to convert these optical depths of the 100 rays into a transit spectrum. First, \texttt{sunbather} adds 50 more rays between $y=0$ and $y=R_p$ with $\tau_y(\nu)=\infty$ to take into account the opaque planet core. Each ray $i$ then corresponds to a projected annulus on the stellar disk (the $y-z$-plane) of area $S_i = \pi (y_{i+1} - y_i)^2$. However, the background stellar flux varies across such an annulus, based on the stellar limb darkening and the projected planet position (determined by its impact parameter $b$ and orbital phase $\phi$). Each annulus is thus divided into 500 cells distributed spherically along the annulus. In each of these cells, the stellar intensity $I(\nu)$ is evaluated according to the (frequency-dependent) quadratic limb darkening law. Taking the average of the 500 cells results in an average intensity $\bar{I}_i(\nu)$ per ray. We then calculate the average intensity of the complete stellar disk $\bar{I}_*(\nu)$ by sampling it in 1000 annuli. Finally, the total transit spectrum can be calculated, defined as the ratio of the flux in-transit $F_{in}(\nu)$ and the flux out-of-transit $F_{out}(\nu)$:
\begin{equation}
    \frac{F_{in}(\nu)}{F_{out}(\nu)} = 1 - \sum_i  \frac{\bar{I}_i(\nu) \cdot S_i}{\bar{I}_*(\nu) \cdot \pi R_*^2} \Big[ 1- \mathrm{exp}(-\tau_{y_i} (\nu)) \Big],
\end{equation}
where $R_*$ is the stellar radius.

Having discussed the main modules of \texttt{sunbather}, we provide a clarification of how the planet radius $R_p$ is defined. The isothermal Parker wind keeps the mass-loss rate constant at all radii, from a singular point at $r \to 0$ until $r \to \infty$. In reality, the planetary outflows are being launched at some finite altitude between the lower atmosphere, which is in hydrostatic equilibrium, and the upper atmosphere, which is well described by a Parker wind profile. Our models do not capture this transition, and so the Parker wind profile is truncated at a lower boundary of $1 R_p$. In calculating the transmission spectrum, we assume that the planet is fully opaque at all wavelengths for $r < R_p$. 

\section{Escaping atmospheres at supersolar metallicity} \label{sec:metallicity}
We employ \texttt{sunbather} to investigate how atmospheric escape depends on the composition. The dependence on the helium abundance in hydrogen/helium outflows has been studied before, and the degeneracy with the extreme UV flux of the host star is well known \citep[][]{lampon_modelling_2021,  khodachenko_simulation_2021, czesla_h_2022, fossati_gaps_2022, xing_mass_2023, biassoni_self-consistent_2024, allan_evolution_2024}. Even though atmospheric escape models that incorporate metals have been employed in the literature \citep[e.g.,][]{salz_simulating_2016, shaikhislamov_3d_2020, huang_hydrodynamic_2023}, a systematic exploration of the influence of metals has so far only been performed by \citet{zhang_escaping_2022} and \citet{kubyshkina_precise_2024}. 

Studying how metals impact atmospheric mass-loss is especially important in the context of small gaseous planets. In the solar system, a clear anticorrelation between planet mass and atmospheric metallicity has been observed \citep{guillot_giant_2022}. However, establishing a similar relationship for exoplanets proves more challenging. While there is evidence supporting an inverse correlation between planet mass and bulk metallicity \citep{thorngren_massmetallicity_2016}, it is not yet clear how this relates to the atmospheric metallicity \citep[e.g.,][]{welbanks_massmetallicity_2019, bean_high_2023}.

In this section, we investigate how the atmospheric bulk properties such as density, velocity and temperature respond to an increasing metal content. We also study how the changing atmospheric structure and abundances in turn affect the transit spectrum. A front-to-back reproduction script for the results of this section is available on Zenodo$^{\ref{fn:zenodo}}$. The simulations presented here neglected the stellar tidal gravity term of Eq. \ref{eq:momentum_cons}, but we checked that the results are qualitatively the same as simulations including it. We simulate a generic hot-Neptune planet with $M_p=0.1$~$M_J$, $R_p=0.5$~$R_J$, orbiting a Sun analog at $a=0.05$~AU. The solar spectrum that we use is a combination of different datasets. The 0.5-189.5~nm wavelength region consists of data from the SEE instrument aboard the TIMED spacecraft, which has been measuring the full solar disk irradiance since 2002 \citep{woods_solar_2005}. We use the Level 3 data product, which constitutes one-day average spectra with flares removed. The data we use is from January 1st 2016. This date is in-between the maximum and minimum of the 11-year solar cycle and was thus assumed to represent a somewhat average spectrum. The 190-310~nm and 310-2412~nm wavelength regions consist of data from the SOLSTICE and SIM instruments, respectively, aboard the SORCE spacecraft that also continuously monitored the solar spectrum \citep{rottman_sorce_2005}. These components are again daily averages at January 1st 2016. We finally extrapolate the red end of the spectrum with a Rayleigh-Jeans power-law until 10,000~nm. The spectrum is shown in Fig. A.1 of \citealt{linssen_expanding_2023}. 

We simulate the planet using six different atmospheric compositions: a pure hydrogen/helium mixture in the ratio 10:1 by number, a complete solar composition including metals, and three, ten, 30, and 50 times solar metallicity. A scaled solar metallicity $Z$ in this case means we scale the metal/hydrogen abundance ratio relative to the solar ratio as
\begin{equation} \label{eq:metallicity}
Z = \Bigg( \frac{X_m}{X_H} \Bigg) \; \Bigg/ \; \Bigg( \frac{X_{m, \odot}}{X_{H,\odot}} \Bigg),
\end{equation}
where $m$ runs over every metal element. With this definition, $Z$ can take any value, but the increase in metal abundance becomes nonlinear at high $Z$. To illustrate, while the solar carbon abundance is $X_C=2.225 \times 10^{-4}$, at 50 times solar metallicity, the abundance increases to $X_C=1.063 \times 10^{-2}$, which is only a factor $\sim$48 increase. The carbon abundance converges towards $X_C=0.233$ as $Z \to \infty$.

\subsection{Assessing the validity of a constant composition} \label{sec:crossover_mass}
As explained in Sect. \ref{sec:code}, \texttt{sunbather} does not treat mass fractionation but instead has a constant composition with altitude. To clarify, the ionization fractions and energy level populations of each element do change as a function of radius, but the overall abundance of the element does not. We can assess the validity of the assumption of constant composition in the high-metallicity outflows by using the concept of the crossover mass, as introduced by \citet{hunten_mass_1987}. In an atmosphere consisting of two constituents, the crossover mass is defined as
\begin{equation} \label{eq:crossover_mass}
    m_c = m_1 + \frac{k T F_1^0}{g_0 X_1 b},
\end{equation}
where $m_1$ is the mass of the first constituent, $k$ is the Boltzmann constant, $T$ is the temperature, $F_1^0$ is the number outflow flux of the first constituent, $g_0=GM_p/R_p^2$ is the planet surface gravitational acceleration, $X_1$ is the abundance of the first constituent, and $b$ is the diffusion parameter, given by \citep{cussler_diffusion_2009}:
\begin{equation} \label{eq:diffusion_parameter}
    b = n \cdot \frac{A T^{3/2}}{P \sigma_{12}^{2} \Omega} \sqrt{\frac{1}{M_1} + \frac{1}{M_2}},
\end{equation}
where $n$ is the number density, $P$ is the pressure, $\sigma_{12}=(\sigma_1+\sigma_2)/2$ is the average collision diameter of the two constituents, $\Omega$ is a dimensionless constant of order unity that we take to be 1 here \citep{hirschfelder_molecular_1967}, $M_1$ and $M_2$ are the molar masses of the two constituents, and
\begin{equation}
    A = \frac{3}{8} k^{3/2} \sqrt{\frac{N_A}{2\pi}},
\end{equation}
where $N_A$ is the Avogadro constant. If we take atomic hydrogen as constituent 1, assuming it drives the outflow and drags along the second constituent (a metal species), the crossover mass can be understood as the particle mass above which the metal species is no longer effectively dragged along. In other words, if $m_2<<m_c$, the metal species is dragged along efficiently, resulting in a constant composition with altitude. However, as $m_2$ approaches $m_c$, the metal escape flux approaches 0. 

We can thus check our model assumption of efficient metal drag by estimating the crossover mass at $r=R_p$. We do this for the model that requires the most extreme metal drag, namely the 50 times solar metallicity case with Parker wind parameters $T_0=5000$~K and $\dot{M}=10^{9.9}$~g~s$^{-1}$ (see Sect. \ref{sec:bulk_structure} for the choice of parameters). With such a composition, $X_1=0.87$ and the hydrogen number flux is $F_1^0=1.3 \times 10^{13}$~s$^{-1}$. In principle, our atmosphere is a mixture of many different constituents, but Eq. \ref{eq:crossover_mass} is derived under the assumption of only two constituents. We take the second constituent to be atomic iron, with $M_2=55.845$~g~mol$^{-1}$. The collision diameters are not readily available, and so we assume $\sigma_{12}=2.58$~Å (similar to the value of various molecular gases, see \citealt{hirschfelder_molecular_1967}). Using the temperature, pressure and number density of the Parker wind profile at $R_p$, we find $b=1.44 \times 10^{20}$~cm$^{-1}$~s$^{-1}$ and $m_c=43$~amu. This means that the assumption of efficient metal drag and a constant composition with altitude is questionable for the heaviest metal species in this particular simulation (zinc is the most massive particle included, with a a mass of 65~amu). The simulations at lower metallicities have higher crossover masses ($m_c = 127$~amu at 30 times solar metallicity) and metal drag can thus be assumed to be efficient in them. Even for the 50 times solar metallicity case however, our calculation used hydrogen and iron in atomic form as the two constituents. For ions, the collision diameter is much stronger \citep{hunten_mass_1987}, and consequently the crossover mass is much larger. The \texttt{Cloudy} simulation shows that both hydrogen and iron (and other metal species for that matter) are indeed mostly ionized already at $r=R_p$. If we re-calculate the crossover mass for an ionized outflow using the more generalized description given in Eq. 33 of \citet{xing_mass_2023}, we find $m_c\approx 5 \times 10^5$~amu. Hence, we consider the assumption of efficient metal drag to be reasonable for the 50 times solar metallicity simulations as well.

\subsection{Effect of metallicity on the outflow structure} \label{sec:bulk_structure}
\subsubsection{The model parameters} \label{sec:model_params}
When comparing models with different metallicities, we have to choose appropriate values for the model free parameters $T_0$ and $\dot{M}$. We first explore how the atmospheric structure changes when using the same mass-loss rate of $\dot{M}=10^{11}$~g~s$^{-1}$ in each model. To find a suitable $T_0$, we perform a model self-consistency check. This means we run multiple models with different values for $T_0$. For each, we calculate the average of the nonisothermal temperature profile between $r=R_p$ and $r=2R_p$ and compare it to $T_0$. We select the model where the difference is the smallest, because the nonisothermal temperature profile is then most consistent with the density and velocity profiles which were derived under the isothermal approximation. We refer the interested reader to \citet{linssen_constraining_2022} and \citet{linssen_expanding_2023} for a more in-depth discussion on assessing the self-consistency of a Parker wind model. We find self-consistent values for $T_0$ of 5100, 5000, 5000, 4900, 4700, 4100~K for the six models ordered by ascending metallicity.

   \begin{figure*}
   \centering
   \includegraphics[width=\hsize]{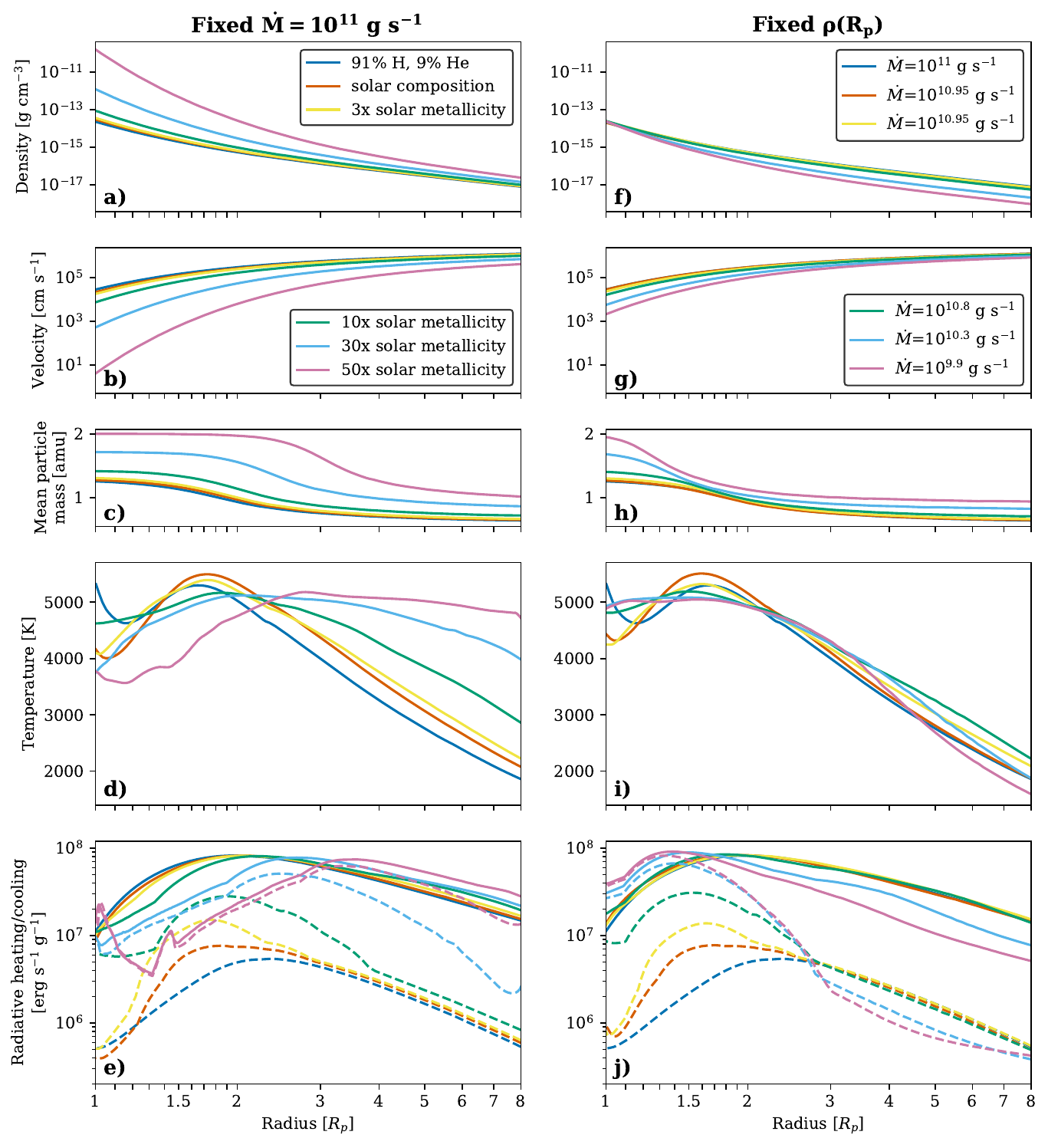}
      \caption{Atmospheric profiles for a generic hot-Neptune planet at different metallicities. The left column shows a set of models with a fixed mass-loss rate. The right column shows another set of models with a fixed density at the planet radius, leading to different mass-loss rates at different metallicities. For every model, the value of $T_0$ is chosen such that it is consistent with the mean of the nonisothermal temperature profile of panel {\bf (d)} or {\bf (i)} between 1$R_p$ and 2$R_p$. Those values for $T_0$ can be found in the text in Sect. \ref{sec:model_params} and Sect. \ref{sec:z_masslossrate}. In panels {\bf (e)} and {\bf (j)}, solid lines show the radiative heating rates and dashed lines show the radiative cooling rates.}
         \label{fig:hotneptune_structure}
   \end{figure*}
   
\subsubsection{Effect of metallicity on the hydrodynamics} \label{sec:z_hydrodynamics}
The left column of Fig. \ref{fig:hotneptune_structure} shows the bulk atmospheric properties of the six models. Panels a, b, and c show the density, velocity and mean particle mass profiles, respectively, which are calculated under the assumption of an isothermal Parker wind. Unsurprisingly, the mean particle mass increases with the metallicity of the atmosphere. This difference in mean particle mass then propagates into different density and velocity profiles through the sound speed (Eq. \ref{eq:sound_speed}). As the sound speed becomes lower, the outflow velocity decreases and depends more steeply on radius. Since the mass-loss rate $\dot{M}=4\pi r^2 \rho(r) v(r)$ is kept constant, the density shows the reverse behaviour and increases while also depending more steeply on radius. These changes only become substantial at metallicities higher than three times solar. The conclusion that the mean particle mass is the main driver of structural changes is in agreement with the findings of \citet{kubyshkina_precise_2024}, who compare a hydrogen, a hydrogen/helium, and a solar composition outflow.

\subsubsection{Effect of metallicity on the temperature} \label{sec:z_temperature}
Panel d shows the {nonisothermal} temperature profiles that we obtain by solving Eq. \ref{eq:energy_cons}. We generally observe a radial flattening of the temperature profile with increasing metallicity. We can understand this by considering the ``thermostat temperature'', which is the temperature that the gas would have in radiative equilibrium (in the absence of the expansion and advection terms). The thermostat temperature has a rather weak dependence on density (and hence atmospheric radius) and is $\sim$10,000~K for the pure hydrogen/helium atmospheric density profile of panel d \citep[agreeing with the findings of][]{murray-clay_atmospheric_2009}. Due to metal cooling, the thermostat temperature steadily decreases towards $\sim$5000~K at 50 times solar metallicity. For the pure hydrogen/helium and lower metallicity atmospheres, the thermostat temperature is not actually reached because of expansion cooling, and we instead find a temperature profile that decreases with radius. On the other hand, for the high metallicity atmospheres, the expansion cooling rates are much lower because of the lower outflow velocities. The high metallicity outflows therefore do reach their thermostat temperature, and we find a temperature profile that is $\sim$5000~K with little dependence on radius. This explanation is supported by the radiative heating and cooling rates shown in panel e. For the lower metallicity models, the radiative heating rate is roughly an order of magnitude higher than the radiative cooling rate. The radiative heating is instead balanced by expansion cooling. For the 50 times solar metallicity model, we see that radiative heating and cooling balance each other throughout the radial domain. The models at ten and 30 times solar metallicity are (almost) in radiative equilibrium only at lower radii, and this is indeed where their temperature profiles are relatively flat. We can thus comment on the thermal properties of the gas by looking at the difference between the radiative heating and cooling rates. At high metallicity, thermal balance between radiative processes indicates that chemistry plays a dominant part in governing the outflow. At lower metallicity, however, the onset of expansion cooling highlights the relevance of dynamics in affecting the atmosphere, especially at higher radii.

We also note that it is somewhat coincidental that the low metallicity models have a temperature of $\sim$5000~K at $r \lesssim 2R_p$, which is similar to the high metallicity thermostat temperature. This is because expansion cooling dominates throughout the atmosphere of the hot-Neptune planet that we simulate here. As identified in for example \citet{salz_simulating_2016,caldiroli_irradiation-driven_2021,linssen_constraining_2022}, a higher gravity planet like a hot Jupiter is in radiative equilibrium at lower radii, and would therefore reach a temperature close to the thermostat value of $\sim$10,000~K.

\subsubsection{Effect of metallicity on the mass-loss rate} \label{sec:z_masslossrate}
It is interesting to examine how the mass-loss rate of the planet changes as a function of atmospheric metallicity. Unfortunately, we cannot easily answer this question, since the mass-loss rate is assumed, rather than predicted, in our model. However, under the assumption that the density at the planet radius stays the same irrespective of the outflow metallicity, we can adjust the mass-loss rate of the models with different compositions in such a way that we respect that density. This assumption can be debated, as modeling efforts have shown that metallicity affects the conditions at the photosphere \citep[e.g.,][]{molliere_model_2015, drummond_effect_2018}. However, we are only aiming to do a simple exploration of the problem here. With this exercise, we are effectively imposing a boundary condition on $\rho(R_p)$ that our isothermal Parker wind model normally does not have. Other atmospheric escape models that treat the hydrodynamics and thermal balance self-consistently also have such a boundary condition. In those models, the chosen boundary conditions can affect the resulting mass-loss rate ranging from a few tens of percent up to a factor of a few, depending on the planet considered \citep[e.g.,][]{salz_simulating_2016, kubyshkina_precise_2024}. We should thus take into account that changes to the mass-loss rate at that level could be attributed to the boundary condition rather than metallicity.

We keep the Parker wind parameters of the pure hydrogen/helium composition the same as before, that is, a self-consistent temperature of $T_0=5100$~K and a mass-loss rate of $\dot{M}=10^{11}$~g~s$^{-1}$. For the simulations with different metallicity, we then find the mass-loss rate for which $\rho(R_p)$ is the same as that of the hydrogen/helium profile, while using a self-consistent $T_0$. This yields the following values:
$T_0=5100$~K and $\dot{M}=10^{10.95}$~g~s$^{-1}$ for the model with a solar composition, 
$T_0=5100$~K and $\dot{M}=10^{10.95}$~g~s$^{-1}$ for the model with three times solar metallicity, 
$T_0=5100$~K and $\dot{M}=10^{10.8}$~g~s$^{-1}$ for the model with ten times solar metallicity, 
$T_0=5000$~K and $\dot{M}=10^{10.3}$~g~s$^{-1}$ for the model with 30 times solar metallicity, 
and $T_0=5000$~K and $\dot{M}=10^{9.9}$~g~s$^{-1}$ for the model with 50 times solar metallicity.
Hence, for a fixed density $\rho(R_p)$, the mass-loss rate is roughly an order of magnitude lower at high metallicity than at solar metallicity. 

The right column of Fig. \ref{fig:hotneptune_structure} presents the atmospheric structure of these models. Since the mass-loss rates of the lower metallicity models are close to $10^{11}$~g~s$^{-1}$, their atmospheric structures are almost the same as in the left column of Fig. \ref{fig:hotneptune_structure}. The differences are larger for the higher metallicity models. The temperature profiles shown in panel i are now all remarkably similar. Like in Sect. \ref{sec:z_temperature}, we can explain this with an argument based on the expansion cooling rate. The velocity and hence expansion cooling rate of the higher metallicity models is relatively low at radii $r\lesssim 2R_p$. As panel j confirms, the result is that these models are in radiative equilibrium at their thermostat temperatures of 5000~K in this region of the atmosphere. Contrary to the previous case at fixed $\dot{M}=10^{11}$~g~s$^{-1}$, their velocities are now close to the velocities of the solar composition model at radii $r \gtrsim 2 R_p$. Due to expansion cooling, their temperature profiles are therefore not flat, but decrease with radius.

\subsubsection{Radiative heating and cooling by metals} \label{sec:z_radheatcool}
\begin{figure*}
   \centering
   \includegraphics[width=\linewidth]{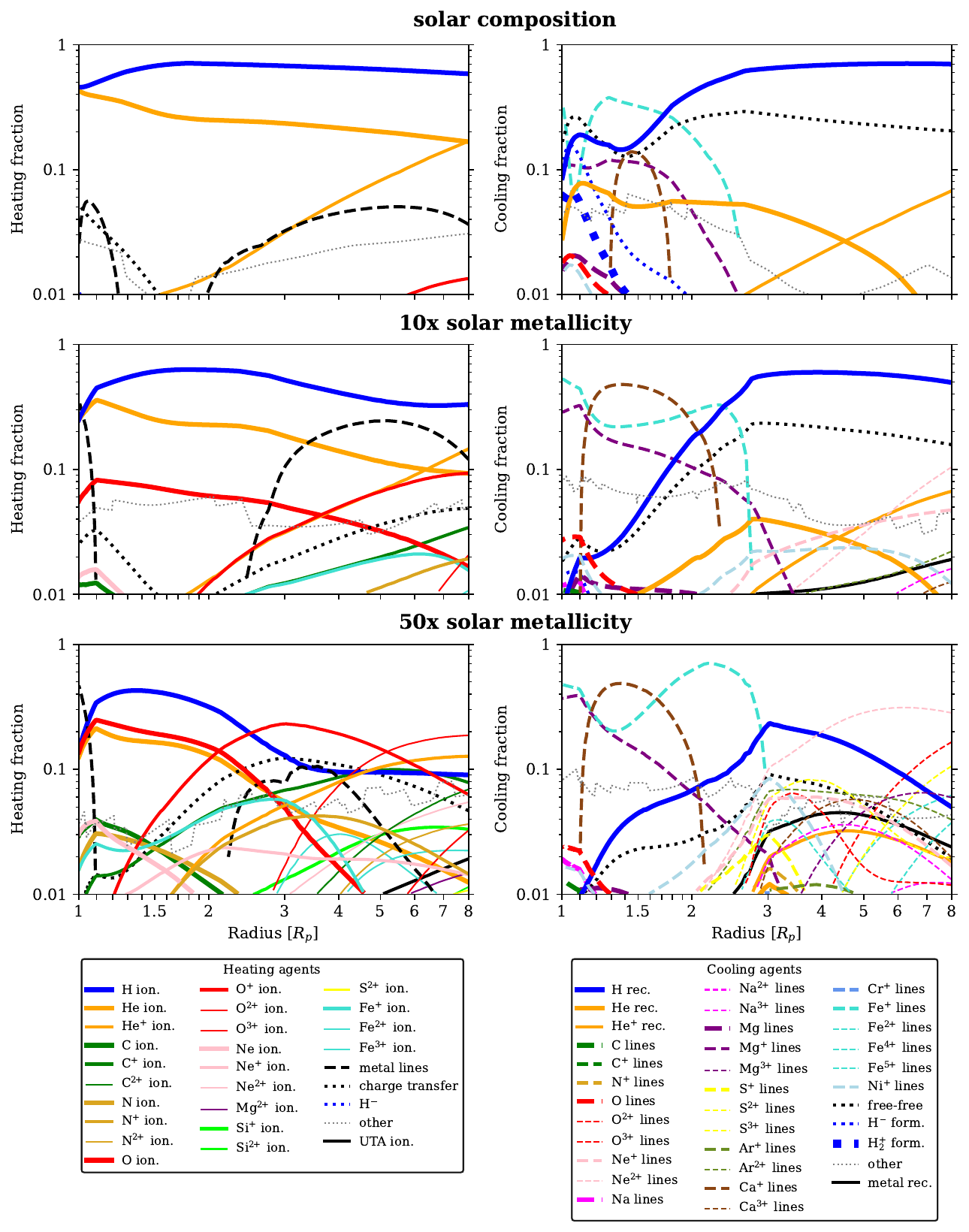}
      \caption{Radiative heating and cooling agents as given by \texttt{Cloudy} for models with a fixed density at the planet radius (right column of Fig. \ref{fig:hotneptune_structure}). The rates are expressed as a fraction of the total radiative heating or cooling rate. The grey dotted line labeled ``other'' is the sum of all other agents that are not plotted as they do not exceed a fraction of 0.01. For some processes such as heating by ``metal lines'', we unfortunately cannot obtain more details from the \texttt{Cloudy} output about the specific elements responsible. Please note the log-scale when interpreting the relative importance of the different agents.} 
         \label{fig:hotneptune_heatcool_agents}
   \end{figure*}

We take a closer look at the different heating and cooling processes that make up the total radiation rates. For the simulations with a fixed density at $R_p$ (right column of Fig. \ref{fig:hotneptune_structure}), Fig. \ref{fig:hotneptune_heatcool_agents} shows the contribution of each process, as given by \texttt{Cloudy}. We present these as a fraction of the total radiative heating or cooling rate. At solar composition, heating is dominated by hydrogen ionization, but helium ionization contributes significantly, especially at smaller radii. Metals only slightly affect the heating through line heating, which makes up $\lesssim5$\%. Cooling is dominated by hydrogen recombination and free-free emission at radii $r\gtrsim 2R_p$. At smaller radii, line cooling from iron, magnesium, and calcium ions becomes important and comparable to the hydrogen recombination cooling rate. Interestingly, cooling through hydrogen Lyman-$\alpha$ emission seems to be unimportant, contrary to the findings of other works both employing \texttt{Cloudy} \citep[e.g.,][]{salz_simulating_2016, zhang_escaping_2022, kubyshkina_precise_2024} and not employing \texttt{Cloudy} \citep[e.g.,][]{murray-clay_atmospheric_2009, allan_evolution_2024}. It is not immediately clear to us why this is the case. As we go to ten times solar metallicity, heating is still mostly dominated by hydrogen and helium ionization, but metal line heating and oxygen ionization is becoming more important. Cooling below $\lesssim 2 R_p$ is now completely dominated by iron, magnesium, and calcium line cooling. At 50 times solar metallicity, heating occurs through wealth of different agents. At smaller radii, hydrogen, helium, and carbon ionization still dominate, but above $2 R_p$, ionization of various metal atoms and ions all contribute a few~\% to 20~\% of the heating. The cooling rate qualitatively shows roughly the same behaviour as at ten times solar metallicity. Iron, magnesium, and calcium line cooling dominate at smaller radii, while at larger radii cooling occurs through the lines of various metal ions, the strongest being Ne$^{2+}$. These results show that metals are important in the radiative budget of the planet already at solar abundance through their line cooling. At enhanced metallicities of approximately ten times solar and above, their ionization and line heating also become important.

Since Fig. \ref{fig:hotneptune_heatcool_agents} shows the fractional contribution of each heating and cooling agent, we need to compare with Fig. \ref{fig:hotneptune_structure}j to get a sense for the absolute change in the heating and cooling rates of the different agents. This reveals that at radii $\gtrsim 1.5 R_p$, the lower hydrogen and helium ionization heating fractions at high metallicity indeed also correspond to a lower radiative heating rate. In other words, in those regions, the decrease in hydrogen/helium heating is not replaced by similar amounts of metal ionization heating. At radii $\lesssim 1.5 R_p$ however, the radiative heating rate does increase with metallicity due to O$^+$ ionization. For the cooling rates, we see similar behaviour, but now at a dividing radius of $\sim 3 R_p$. At radii $\lesssim 3 R_p$, the loss of hydrogen recombination cooling is not counterbalanced by metal cooling. However, at radii $\lesssim 3R_p$, the onset of iron, magnesium, and calcium line cooling does lead to an overall much higher cooling rate. 

\subsection{Effect of metallicity on the transit spectrum} \label{sec:z_spectrum}
\begin{figure}
   \centering
   \includegraphics[width=\hsize]{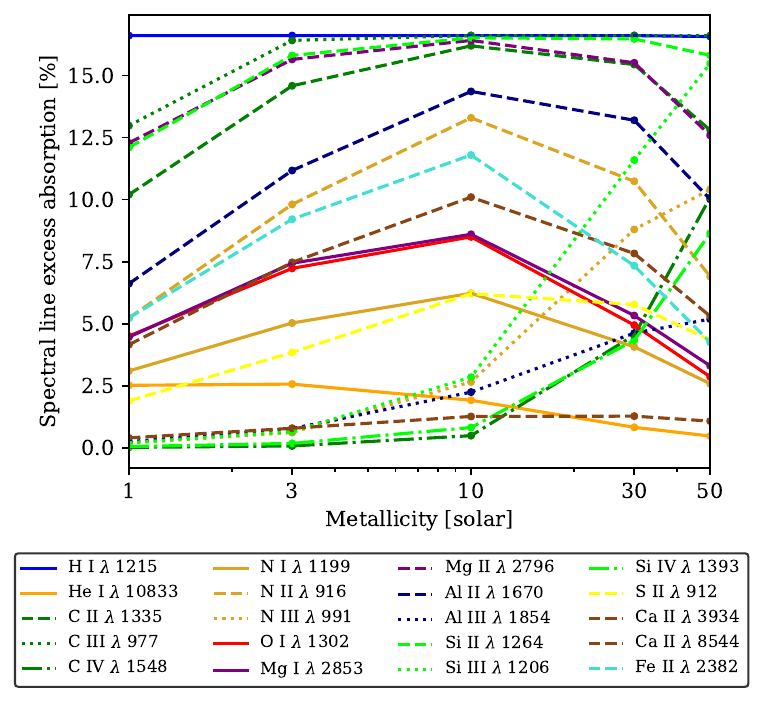}
      \caption{Depths of the 18 strongest lines in the transit spectra, as well as the metastable helium and calcium infrared triplet lines. The results are shown for the models with a fixed density at the planet radius (right column of Fig. \ref{fig:hotneptune_structure}). Our computational radial domain extends to 8$R_p$, corresponding to a maximum absorption of 16.9\%, which is reached by various spectral lines. Lines originating from metal atoms (solid) and singly charged ions (dashed) tend to become stronger up to ten times solar metallicity, and then become weaker. Lines originating from doubly or higher charged metal ions (dotted and dash-dotted) keep growing stronger as the metallicity increases.}
         \label{fig:hotneptune_tds}
   \end{figure}

To investigate the effect of the outflow metallicity on the transit spectrum, we calculated the transit depths of various atmospheric escape tracers. We first calculated the transit depths of all spectral lines identified and listed in Table C.1 of \citet{linssen_expanding_2023}. From these, we selected the 18 strongest lines and supplemented this with the metastable helium line and calcium infrared triplet, to obtain a sample of 20 spectral lines. Fig. \ref{fig:hotneptune_tds} presents the transit depths of these spectral lines as a function of outflow metallicity, for the simulations with a fixed photospheric density (right column of Fig. \ref{fig:hotneptune_structure}). The spectra were made at mid-transit and without stellar center-to-limb variations. 

The hydrogen Lyman-$\alpha$ line appears to have a transit depth independent of metallicity. This is an artifact of our atmospheric radial grid, which extends to 8$R_p$ so that the maximum transit depth is $(8R_p/R_*)^2=16.9\%$. Even though there is certainly less absorption in the Lyman-$\alpha$ line at high metallicity, the line is still optically thick and hence its transit depth does not decrease. A model that extends to larger radii (and takes into account interactions with the stellar environment) would be needed to properly explore the dependence of the Lyman-$\alpha$ line on the outflow metallicity. Since the metastable helium line forms at smaller radii, it is seen to become weaker at high metallicity. This is not due to the decrease in helium abundance, which is only $\sim 4\%$, but due to structural changes in the outflow. The dominant factor is that the total density (hence also helium density) decreases because the mass-loss rate gets lower. At the same time, Fig. \ref{fig:hotneptune_structure} shows that the temperature decreases at radii $r \lesssim 2 R_p$ and increases at radii $r \gtrsim 2 R_p$ at higher metallicity, leading to an increase and decrease in the fraction of helium atoms in the metastable state, respectively \citep[see Fig. 1 of][]{biassoni_self-consistent_2024}. The combined result turns out to be a weaker helium line.

For all the other lines that originate from metals, we notice an interesting response to an increasing metallicity. Spectral lines originating from atomic or singly ionized metals become stronger until ten to 30 times solar metallicity and then become weaker again at 50 times solar metallicity. In contrast, spectral lines originating from doubly or higher ionized metals keep growing stronger until 50 times solar metallicity. This behavior is the result of a decreasing number density of atoms and singly charged ions on the one hand and an increasing number density of doubly and higher charged ions on the other hand. We can understand why the ionization fractions change in this way as follows. If we temporarily consider a total density profile that is kept constant, increasing the metal content will lead to a lower free electron density. This is due to the fact that a metal particle generates fewer free electrons per unit mass than hydrogen or helium - assuming the metal particle is not completely ionized, which is indeed not the case in the conditions of an upper exoplanet atmosphere. The lower electron density leads to a lower recombination rate of the ions, as the radiative recombination rate depends linearly on the electron density \citep{rybicki_radiative_1991} and the three-body recombination rate depends on the electron density squared. 
The photoionization rate stays roughly the same, and as a consequence the degree of ionization increases. At 50 times solar metallicity, this effect is so strong that we get a lower number density of metal atoms and singly charged ions, even though the overall metal abundance is higher than in the 30 times solar metallicity case. Vice versa, the number density of doubly and higher charged ions grows faster than the increase in overall metal abundance. Finally, we can consider the case where the total density profile is not kept constant. As the top-right panel of Fig. \ref{fig:hotneptune_structure} shows, the total density in fact decreases with metallicity, which adds to the weakening of the lines.

\section{Constraining mass-loss rates from spectral line observations} \label{sec:constraining}
In this section, we demonstrate one of the main use cases of \texttt{sunbather}: fitting an observed transit spectrum with a grid of models in order to constrain the planet's mass-loss rate. Measuring the mass-loss rate is especially interesting as it is a fundamental parameter in the atmospheric evolution. Obtaining accurate mass-loss rates for individual planets will help to reveal the role of atmospheric escape in shaping the sub-Jovian desert and radius valley \citep[e.g.,][]{owen_photoevaporation_2018, owen_kepler_2013}. Constraining the mass-loss rate with \texttt{sunbather} is enabled by the fact that it is a free parameter in the Parker wind description, and can thus be adjusted until the model fits the observations. With this in mind, many works have indeed used an isothermal Parker wind model to fit observations of the metastable helium line (see references in the introduction). In most cases, the fit is performed only for $T_0$ and $\dot{M}$ while considering a hydrogen/helium outflow with some assumed helium abundance. However, some of these works have additionally fit for the helium abundance \citep[e.g.,][]{lampon_modelling_2020, dos_santos_p-winds_2022}, but it is typically degenerate with other model parameters. In Sect. \ref{sec:mock_retrieval}, we start by demonstrating the method on a mock spectrum, which serves as a simple example. We then move on to an observed spectrum of a real planet in Sect. \ref{sec:constraints_z0}, which serves as a more complicated example. In Sect. \ref{sec:constraints_z100}, we explore the influence of the assumed metal abundance on the fit, and show that even though we keep the hydrogen/helium ratio the same, the presence of metals in the outflow still affects the mass-loss rate that is inferred from helium observations. A front-to-back reproduction script for the results of this section is available on Zenodo$^{\ref{fn:zenodo}}$, and it can easily be modified to interpret other observations.

\subsection{Fitting a mock spectrum} \label{sec:mock_retrieval}
\begin{figure}[h!]
   \centering
   \includegraphics[width=\linewidth]{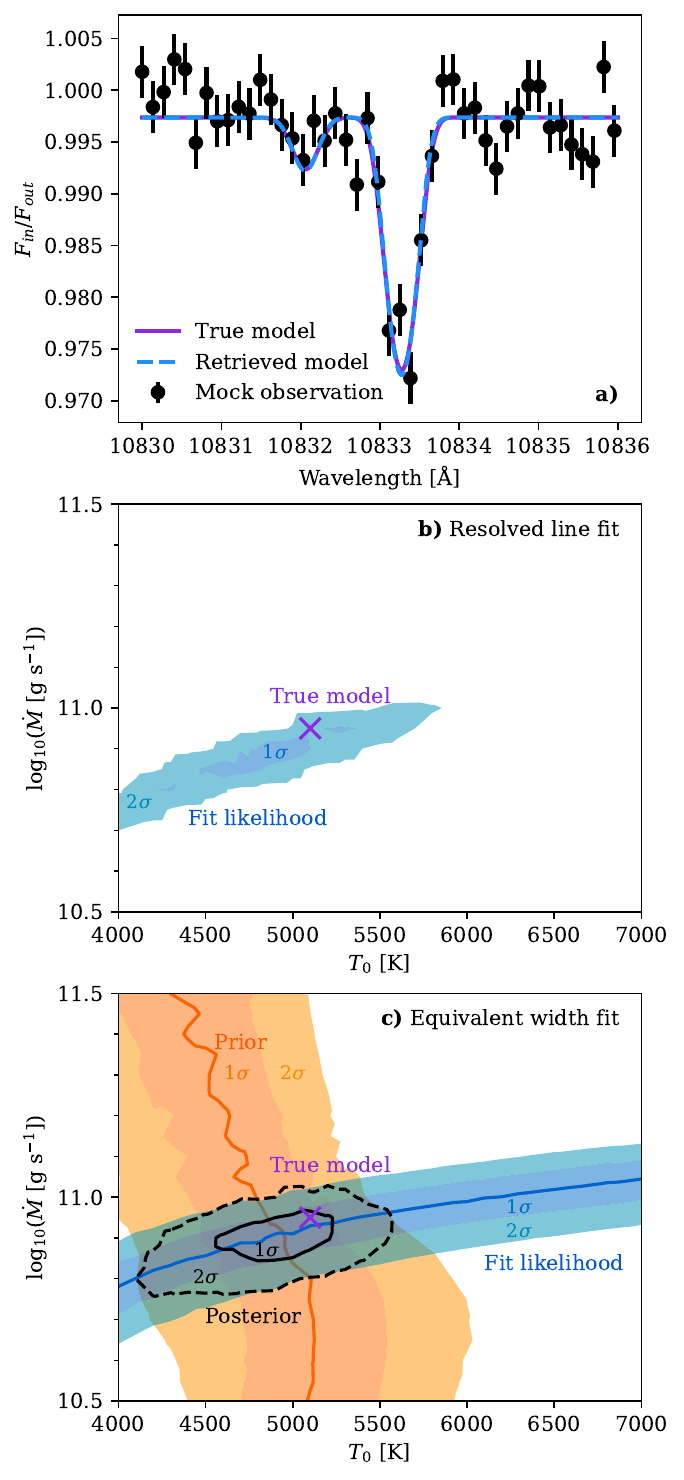}
      \caption{Demonstration of a retrieval of the mass-loss rate and temperature from the helium spectrum of the generic hot-Neptune introduced in Sect. \ref{sec:metallicity}. \textbf{a)} True model metastable helium spectrum and the mock observed spectrum. The best retrieved model from the resolved line fit is also shown. \textbf{b)} Results of the fit to the spectrally resolved line, without using a prior. {\bf c)} Results of the fit to the EW of the line. As visible from the blue contours, the fit offers no tight constraints on its own. We thus place a normal prior on the parameter space based on the model temperature self-consistency, shown in orange. The posterior distribution shown in black then yields a precise and accurate constraint.}
         \label{fig:hotNeptune_constraints}
   \end{figure}

We start by generating a fake observed metastable helium spectrum. We take the generic hot-Neptune planet that we introduced in Sect. \ref{sec:metallicity} and choose a solar composition atmosphere, $T_0=5100$~K and $\dot{M}=10^{10.95}$~g~s$^{-1}$. We first use \texttt{sunbather} to calculate the metastable helium spectrum between 10,830 and 10,836~Å, shown as the purple line in Fig. \ref{fig:hotNeptune_constraints}a. We then turn this synthetic spectrum into a mock observed spectrum by introducing noise. We sample the synthetic spectrum at a spectral resolution of 80,000. At each wavelength point, we add random values drawn from a normal distribution with standard deviation 0.0025 to the model $F_{in}/F_{out}$ value. We then also assign an observed errorbar of 0.0025 to the drawn data points. This results in the mock spectrum shown in black in Fig. \ref{fig:hotNeptune_constraints}a.

We now pretend that this is an observed spectrum of the planet and aim to fit for the mass-loss rate and temperature. We fix all other parameters (see Fig. \ref{fig:sunbather_chart} for the parameters that are needed) at the values that we used to create the mock spectrum, thus effectively assuming we know them precisely. For $R_p$, $a$, and $R_*$, this is a reasonable assumption as for real planets they are usually known well enough to not dominate the uncertainties of the analysis. The value of $M_p$ typically has larger uncertainties. To get a sense for when the uncertainty in $M_p$ may become significant, we ran an identical model as that used to generate the mock spectrum, but increased the planet mass by 30\%. The resulting helium spectrum was within $1 \sigma$ of the original spectrum. When increasing the planet mass by 60\%, the difference became more significant at $\sim 3 \sigma$. The final two ``parameters'' that we fix at the real values are the stellar spectral energy distribution (SED) and the planet atmospheric composition, which are typically not known accurately for real planets, but may also significantly influence the results of the analysis. One may thus wish to also explore different SEDs and compositions, in addition to $T_0$ and $\dot{M}$ (we explore this in more detail in Sect. \ref{sec:constraints_z0} for the atmospheric composition). 

We run a grid of models, with temperatures ranging from 4000 to 7000~K in steps of 100~K, and mass-loss rates ranging from 10$^{10.5}$ to 10$^{11.5}$~g~s$^{-1}$ in steps of 0.05~dex (totaling 651 models). Each model calculation consists of first calling the \texttt{construct\_parker.py} routine to calculate the isothermal Parker wind density and velocity profiles (Sect. \ref{sec:construct}), then calling the \texttt{convergeT\_parker.py} routine to calculate the nonisothermal temperature profile (Sect. \ref{sec:convergeT}), and finally using the \texttt{RT.py} module to calculate the metastable helium spectrum (Sect. \ref{sec:RT}).

With all model spectra calculated, we can turn to the statistics. We first resample each model spectrum onto the same wavelength grid as the observed spectrum. Assuming normally distributed errors, we can fit each model to the data using a $\chi^2$-statistic. In a Bayesian framework, these $\chi^2$-values are then related to the likelihood $L$ as $L = \mathrm{exp}(-\chi^2 / 2)$, which we show in Fig. \ref{fig:hotNeptune_constraints}b. One could choose to also use a (nonflat) prior, but we abstain from doing so here since the fit itself is already quite constraining. The purple cross in Fig. \ref{fig:hotNeptune_constraints}b marks the true parameters that were used to generate the mock spectrum. It falls within the 2$\sigma$ contour of the likelihood, confirming that our retrieval has been successful. 

To demonstrate a case in which we have spectrally unresolved data, we now repeat the analysis but fit only the equivalent width (EW) of the line. Observations with the James Webb Space Telescope \citep{dos_santos_observing_2023}, the Hubble Space Telescope \citep{spake_helium_2018}, and the Palomar/WIRC ultranarrowband filter \citep{vissapragada_constraints_2020} would produce such low-resolution or unresolved metastable helium spectra. We integrate our true model spectrum to obtain its EW and then again randomly sample a value around it, obtaining a mock observed EW=$13.5 \pm 1.8$~mÅ (this value is thus not related to the mock resolved spectrum we generated before). We can then simply compare the EW of each model to this observed value, which directly gives the likelihood (shown in blue in Fig. \ref{fig:hotNeptune_constraints}c). As visible from the figure, the fit itself is not very constraining, and we instead find a degeneracy between $T_0$ and $\dot{M}$. 

To mitigate this problem (which is common when fiting Parker wind models to spectrally unresolved data), we follow the approach outlined in \citet{linssen_constraining_2022}. This comes down to now using a nonflat prior to break the degeneracy. We choose a prior that is based on the self-consistency of the atmospheric structure. Specifically, for each Parker wind model we compare the nonisothermal temperature profile $T(r)$ to the isothermal value $T_0$. A large difference between the two indicates that the model is not very self-consistent. To compare the $T(r)$ profile to a single value $T_0$, we need to somehow collapse $T(r)$ into a single value as well. Since we are fitting metastable helium observations, we care most about the self-consistency of the model in the region where this line forms. Therefore, we calculate the mean and standard deviation of $T(r)$ weighted by the metastable helium density, resulting in $T_{He}$ and $\sigma_T$, respectively (Eqs. 4 and 5 of \citealt{linssen_constraining_2022}). Our prior value for the model is then given by a normal distribution centered on $T_{He}$ with standard deviation $\sigma_T$ evaluated at $T_0$. The orange line in Fig. \ref{fig:hotNeptune_constraints}c show models for which $T_{He}=T_0$, while the dark and light orange contours indicate models where $T_0$ falls within 1$\sigma_T$ and 2$\sigma_T$ of $T_{He}$. 

The posterior distribution now follows by multiplying the prior and likelihood and normalizing it. In Fig. \ref{fig:hotNeptune_constraints}c, the solid and dashed black lines indicate the 1$\sigma$ and 2$\sigma$ contours, respectively. We see that the retrieval has been successful, which is in large part due to using the prior. Without it, the fit would have been quite unconstraining for both the temperature and mass-loss rate. We finally note that in the case of this mock spectrum, the temperature self-consistency prior that we used in the equivalent width fit is also consistent with the temperature that we constrained directly from the resolved line fit. As we show in the following section, this is not always the case for real spectra.

\subsection{Fitting the helium spectrum of TOI-2134~b} \label{sec:constraints_z0}
We now turn to a real observed spectrum, which will prove to be more complicated to interpret. We use this example to illustrate how we can use \texttt{sunbather} to derive insights into outflow physics when our current modeling struggles to self-consistently match the data. 

\begin{figure}[h!]
   \centering
   \includegraphics[width=\linewidth]{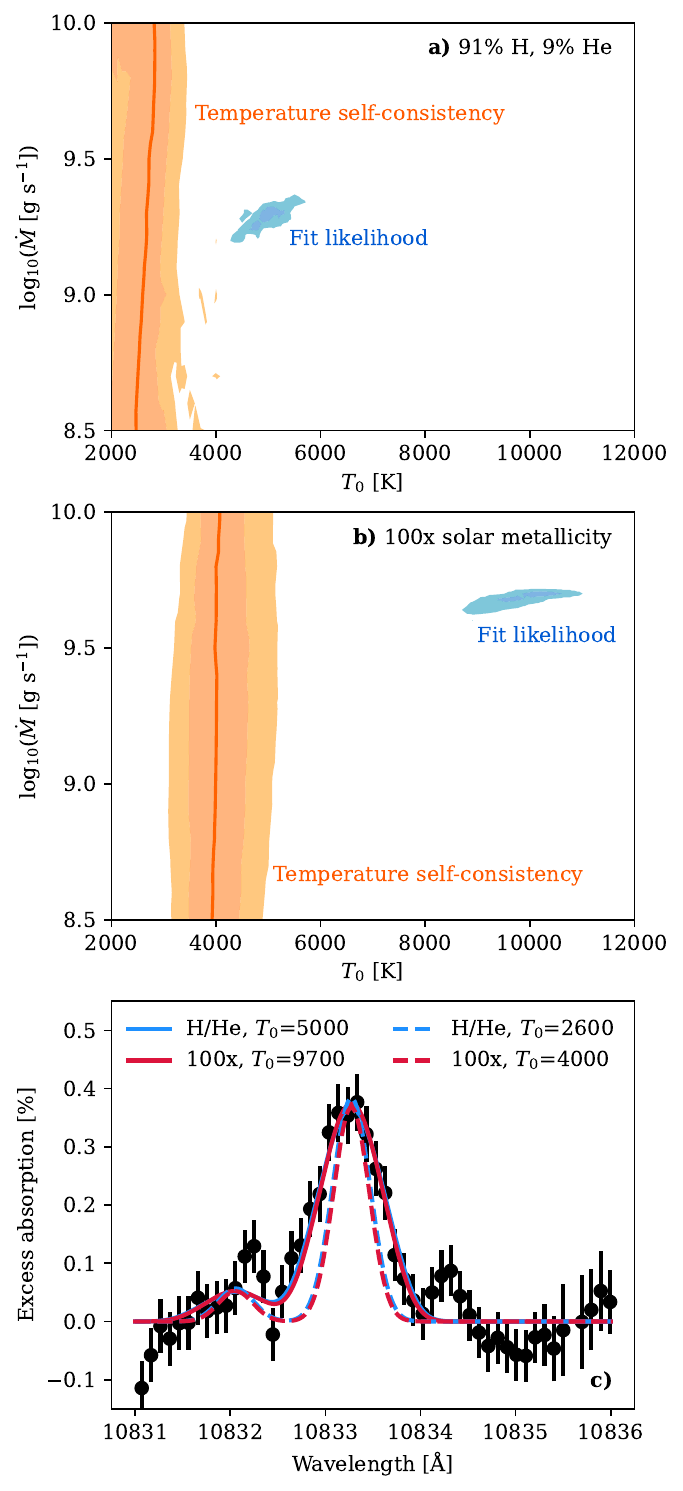}
      \caption{Two fits to the metastable helium data of the mini-Neptune TOI-2134~b from \citet{zhang_outflowing_2023} assuming different atmospheric compositions. In panel {\bf (a)}, we assume a pure hydrogen/helium atmosphere in the solar ratio. In panel {\bf (b)}, we assume a 100 times solar metallicity atmosphere. The colors are the same as in Fig. \ref{fig:hotNeptune_constraints}, but here we do not use the model temperature self-consistency as a prior as we did in Fig. \ref{fig:hotNeptune_constraints}c. {\bf c)} Observations from \citet{zhang_detection_2023} shifted back to the rest-frame helium triplet wavelength (black). Solid lines show the best-fit models from the blue contours in panels (a) and (b). Dashed lines show models with self-consistent $T_0$-values from the orange contours in panels (a) and (b). The self-consistent models produce lines that are too narrow, indicating that an additional line broadening mechanism may be at play.}
         \label{fig:TOI2134b_constraints}
   \end{figure}

We model the helium line of the mini-Neptune TOI-2134~b as observed with Keck/NIRSPEC by \citet{zhang_outflowing_2023}. The original spectrum showed a $\sim$5~km~s$^{-1}$ redshift from the rest-frame helium triplet wavelength. Such offsets can have valuable physical implications \citep[e.g.,][]{nail_effects_2024}, but cannot be reproduced by \texttt{sunbather}, which by virtue of its 1D nature yields lines centered on the rest-frame wavelength. So as to not obtain biased fit results, we first shift the spectrum back to the rest-frame wavelength (shown in Fig. \ref{fig:TOI2134b_constraints}c). We then adopt the system parameters from \citet{zhang_outflowing_2023}: $R_p=0.24$~$R_J$, $M_p=0.0287$~$M_J$, $R_*=0.709$~$R_\odot$, $a=0.078$~AU, $b=0.2$, and use their stellar SED that is reconstructed from XMM-Newton observations. 

We first run models assuming a pure hydrogen/helium composition in the solar ratio of 10:1 by number. The helium abundance is thus $X_{He}=0.091$. We run a grid with $2000<T_0<8000$~K and $10^{8.5}<\dot{M}<10^{10}$~g~s$^{-1}$. The simulations excluded the stellar tidal gravity term of Eq. \ref{eq:momentum_cons}, but we verified that the differences with models including it are minimal for this planet. We calculate the helium spectrum for each model, convolve it to the instrument resolution of $R=32,000$, and fit this to the observed spectrum using a $\chi^2$-statistic. The associated likelihood contours are shown in blue in Fig. \ref{fig:TOI2134b_constraints}a. We find $\dot{M}=10^{9.28^{+0.06}_{-0.04}}$~g~s$^{-1}$ and $T_0=5000 \pm 400$~K and the best-fit model is shown as the solid blue line in Fig. \ref{fig:TOI2134b_constraints}c. \citet{zhang_outflowing_2023} also modeled their own observations using \texttt{p-winds} and found $\dot{M}=10^{9.74 \pm 0.08}$~g~s$^{-1}$ and $T_0=4640\pm 230$~K. The difference with our constraints likely stems from the fact that \texttt{sunbather} calculates a nonisothermal temperature structure. The best-fit models with $T_0\approx 5000$~K have nonisothermal profiles that are quite a bit colder than 5000~K. At those lower temperatures, the metastable helium fraction becomes higher, and a lower mass-loss rate is therefore needed to match the observed line \citep[e.g.,][]{biassoni_self-consistent_2024}. 

We also calculate the model self-consistency in the same way as in Sect. \ref{sec:mock_retrieval}. We find a self-consistent temperature of $T_0 \sim 2600$~K, independent of the mass-loss rate (orange contours in Fig. \ref{fig:TOI2134b_constraints}a). This value is significantly lower and inconsistent with the temperature obtained from the spectral line fit. Because the difference is so large and the line fit itself already yields tight constraints, we do not use the model self-consistency as a prior here. However, it is interesting to discuss the implications and potential causes of the $T_0$-discrepancy. The dashed blue line in Fig. \ref{fig:TOI2134b_constraints}c shows the spectrum of a (self-consistent) model with $T_0=2600$~K and $\dot{M}=10^{8.8}$~g~s$^{-1}$ (the value of the mass-loss rate is manually chosen so that the peak absorption is matched). We see that the $T_0$-discrepancy in practice means that the observed spectral line is broader than expected from our self-consistent models. 

The $T_0$-discrepancy can stem from a variety of factors. It may be partly due to the way in which we assess the self-consistent $T_0$ in the first place. Using a different criterion than the metastable helium density could lead to a different temperature constraint. We can explore this line of thinking a little further by noting that we would obtain the highest possible self-consistent temperature if we were to compare the peak of the nonisothermal temperature profile to $T_0$. With this criterion, we obtain a hard upper limit of $T_0\lesssim 4000$~K, which is still somewhat inconsistent with the temperature preferred by the data. The next possibility to consider is that our models underestimate the temperature of the outflow. This may be due to uncertainties in the stellar SED, or additional heating from, for example, Joule heating \citep{cohen_heating_2024}. If applicable, these additional heat sources might increase the self-consistent temperature of the outflow, leading to larger outflow velocities that broaden the line proportionately. Alternatively,  it is also possible that the thermodynamics are modeled appropriately and the temperature is indeed around 2600~K, but there are additional kinematic components not considered by \texttt{sunbather} that broaden the velocity distribution of the line. Such extra velocity components could arise due to the stellar tidal gravity, the stellar radiation pressure \citep[e.g.,][]{bourrier_3d_2013}, the stellar wind ram pressure \citep[e.g.,][]{macleod_stellar_2022} and its interaction with a planetary magnetic field \citep[e.g.,][]{carolan_effects_2021}, atmospheric turbulence and circulation, and the tidally locked rotation of the planet \citep[e.g.,][]{huang_hydrodynamic_2023}.

Ultimately, observing more spectral lines than only the metastable helium line would help to discriminate between the various hypotheses. Other spectral lines can probe different elements, complementary regions of the atmosphere where the temperature and velocity is different, and have varying natural and thermal line widths, all of which may help us understand the apparent inconsistency \citep{linssen_expanding_2023}.

\subsection{Assuming a 100 times solar metallicity composition} \label{sec:constraints_z100}
We repeat the previous analysis of TOI-2134~b while assuming a 100 times solar metallicity atmosphere. The helium abundance is $X_{He}=0.083$ in this case, which is roughly 10\% lower than in Sect. \ref{sec:constraints_z0}. The results are presented in Fig. \ref{fig:TOI2134b_constraints}b. We find $\dot{M}=10^{9.68 \pm 0.04}$~g~s$^{-1}$ and $T_0=9700 \pm 700$~K. We again also calculate the model self-consistent temperature, which is $T_0 \sim 4000$~K.

We can use the findings of Sect. \ref{sec:metallicity} to explain the changes relative to the pure hydrogen/helium fit. 
As visible from Fig. \ref{fig:TOI2134b_constraints}, the fit at 100 times solar metallicity prefers a much higher $T_0$. A higher $T_0$ leads to higher outflow velocities and it thus compensates for the otherwise lower velocities at high metallicity (see Sect. \ref{sec:z_hydrodynamics}). The velocity must remain roughly the same because it is an important contributor to the spectral line broadening that we are fitting. The fit at 100 times solar metallicity also prefers a higher $\dot{M}$. This is due to the well-known $T_0-\dot{M}$-degeneracy mechanism: at higher $T_0$, the overall density and hence metastable helium density decreases, and a higher mass-loss rate is needed to compensate.
Focusing on the self-consistent temperature, we see it has increased from $T_0 \sim 2600$~K for pure hydrogen/helium, to $T_0\sim4000$~K at 100 times solar metallicity. This is because the temperature profile becomes flatter at higher metallicity and approaches the thermostat temperature (see Sect. \ref{sec:z_temperature}) of $\sim$4000~K. We thus find a self-consistent $T_0$ of that value. We plot the helium spectrum of a model with $\dot{M}=10^{9.1}$~g~s$^{-1}$ in Fig. \ref{fig:TOI2134b_constraints}c (this mass-loss rate is manually chosen to match the peak absorption of the data).

The results of this and the previous section show that the inferred mass-loss rate differs by up to half a dex depending on the assumed outflow metallicity. The helium abundance itself is hardly different, but changes in the outflow structure caused by the high metallicity still affect the metastable helium line. As shown by \citet[][]{lampon_modelling_2021}, changing the assumed helium abundance can furthermore change the inferred mass-loss rate by orders of magnitude. Concluding, the retrieved mass-loss rate is degenerate with the helium abundance, and to a lesser extent, the metallicity. Given that we usually do not know the upper atmospheric composition of a given exoplanet, we again pose that we can achieve better mass-loss-rate constraints by combining observations of multiple spectral lines that trace atmospheric escape.

\section{Summary} \label{sec:summary}
We present \texttt{sunbather}, an open-source$^{\ref{fn:github}}$ code that can be used to model escaping exoplanet atmospheres and their transit spectra. The code calculates the atmospheric density and velocity profiles assuming an isothermal Parker wind using the \texttt{p-winds} code. A refined nonisothermal temperature profile is then calculated using the photoionization code \texttt{Cloudy}. A radiative transfer module finally allows the calculation of the transit spectrum over an arbitrary wavelength range. 

The assets of \texttt{sunbather} are:
\begin{itemize}
    \item It has a highly detailed NLTE treatment of the ionization and excitation state and temperature of the atmosphere through the use of \texttt{Cloudy}. The calculation of a nonisothermal temperature profile should produce transit spectra that are  more accurate than those provided by an isothermal Parker wind model.
    \item It allows an arbitrary composition that includes the 30 lightest elements. Metals are completely accounted for in both the hydrodynamics and thermodynamics. The transit spectrum includes metal lines, and enables joint modeling of multiple spectral lines. 
    \item It is highly suited to the interpretation of spectral line observations through a parametrized description of the atmosphere. The mass-loss rate is a free parameter of the model, allowing one to retrieve it from observations.
    \item It has a reasonably fast runtime of a few minutes to an hour, depending on the composition and density of the atmosphere. This runtime supports parameter space exploration for individual planets and/or studies at the population level (but may require a computer cluster to do so).
\end{itemize}

The limitations of \texttt{sunbather} are:
\begin{itemize}
    \item It is a 1D code, while atmospheric escape is an inherently 3D process. For example, day-to-night variations, magnetic fields, and interaction with a stellar wind can significantly alter the outflow and observational signatures \citep[e.g.,][]{nail_effects_2024, schreyer_using_2024}. 
    \item It keeps the composition of the atmosphere constant with radius. Every element escapes at the same rate, and it thus implicitly assumes efficient drag of heavier elements. This assumption may be invalid, particularly at low mass-loss rates.
    \item It does not treat the hydrodynamics and thermodynamics self-consistently, but instead parametrizes the outflow with a chosen mass-loss rate. As pointed out when describing the assets of the code above, it is this feature that makes  \texttt{sunbather} well suited to retrievals. The downside is that it is less suited to forward modeling and theoretical studies designed to predict the mass-loss rate.
\end{itemize} 

One may prefer to use \texttt{sunbather} over \texttt{p-winds} when fitting spectrally unresolved data and/or metal lines that are not included in \texttt{p-winds}. Conversely, one may prefer to use \texttt{p-winds} when fitting lines included in it over a large parameter space, due to its significantly faster runtime.

We use \texttt{sunbather} to explore the effect of atmospheric metallicity on the escape process. We simulate a generic hot-Neptune planet orbiting the Sun, with different atmospheric compositions ranging from hydrogen/helium-only to 50 times solar metallicity. If we keep the mass-loss rate fixed, a higher metallicity leads to a denser but slower wind, while the temperature profile starts to become isothermal. However, if we instead keep the density at the planet radius fixed, we can gain insight into the effect of metallicity on the mass-loss rate. We then find that a higher metallicity leads to a more tenuous and slower wind, with little effect on the temperature profile. The mass-loss rate decreases by an order of magnitude as we move from solar-like to 50 times solar metallicity. We furthermore find that metals significantly contribute to the radiative cooling rate already at solar abundances. They become important in the radiative heating rate at approximately ten times solar metallicity and above. Regarding the planet's transit spectrum, all metal lines become stronger up to  ten times solar metallicity. At even higher metallicity, lines of atoms and singly charged ions generally become weaker owing to the lower mass-loss rate, while lines of doubly and higher charged ions generally continue to grow stronger. We do not investigate whether all of the described effects of high metallicity also pertain to other types of star--planet systems.

After validating \texttt{sunbather}'s ability to retrieve a known model from a mock spectrum, we used the code to interpret the metastable helium spectrum of mini-Neptune TOI-2134~b as observed by \citet{zhang_outflowing_2023}. If we assume a pure hydrogen/helium composition in the ratio of 10:1 by number, we find a mass-loss rate of $\dot{M}=10^{9.28^{+0.06}_{-0.04}}$~g~s$^{-1}$ and a temperature of $T_0=5000 \pm 400$~K. However, a model self-consistency check indicates a much cooler temperature of $T_0\sim2600$~K, which is hard to reconcile with the data, unless other types of line broadening not accounted for here (e.g., atmospheric turbulence and circulation) play a significant role. If we instead assume a 100 times solar metallicity composition, we find a mass-loss rate of $\dot{M}=10^{9.68 \pm 0.04}$~g~s$^{-1}$ and a temperature of $T_0=9700 \pm 700$~K. This temperature is significantly different from the self-consistent temperature of $T_0\sim4000$~K, which again potentially indicates additional physics. In order to constrain the composition, instead of assuming it, additional observations of other spectral lines would be valuable. Such observations would also help to elucidate the interesting temperature differences that we find.

\begin{acknowledgements}
We thank the anonymous referee for their comments which helped improve this paper. We are grateful to G. Ferland, M. Chatzikos and the rest of the \texttt{Cloudy} team, as well as L. dos Santos and the rest of the \texttt{p-winds} team, for making their codes freely available. We appreciate the helpful discussions with F. Nail and N. Bollemeijer regarding the contents of this manuscript. We thank H. Linssen for designing the \texttt{sunbather} logo. This work has made use of the Snellius Supercomputer operated by SURF. A. Oklop\v{c}i\'{c} gratefully acknowledges support from the Dutch Research Council NWO Veni grant.
\end{acknowledgements}

\FloatBarrier

\bibliographystyle{aa}
\bibliography{library}

\end{document}